\documentclass[prd,aps,preprint,superscriptaddress]{revtex4-1}%
\usepackage{amsmath}
\usepackage{graphicx}
\usepackage{bm}
\usepackage{subcaption} 
\usepackage[colorlinks,allcolors=blue]{hyperref} 
\usepackage{orcidlink}  %

\def\beq{\begin{equation}}
\def\eeq{\end{equation}}

\def\to {\rightarrow}

\begin{document}

\title{Enhancing the sensitivity to FCNC top decays $t\to cH $ and $t\to cS $ in the boosted regime at CLIC }

\author{Shuo Yang\,\orcidlink{0000-0003-0955-3229}}
\email{shuoyang@lnnu.edu.cn}
\affiliation{School of Physics and Electronic Technology, Liaoning Normal University, Dalian, 116029, China}
\affiliation{Center for Theoretical and Experimental High Energy Physics, Liaoning Normal University, Dalian, 116029, China} 

\author{Peng-Bo Zhao\,\orcidlink{0009-0008-8310-701X}}
\email{pengbozhao2002@163.com}
\affiliation{School of Physics and Electronic Technology, Liaoning Normal University, Dalian, 116029, China}
\affiliation{Center for Theoretical and Experimental High Energy Physics, Liaoning Normal University, Dalian, 116029, China}

\author{Ji-Chong Yang\,\orcidlink{0000-0001-6688-5583}}
\email{yangjichong@lnnu.edu.cn}
\affiliation{School of Physics and Electronic Technology, Liaoning Normal University, Dalian, 116029, China}
\affiliation{Center for Theoretical and Experimental High Energy Physics, Liaoning Normal University, Dalian, 116029, China}

\vspace{-5em}  %
\begin{abstract}
The top quark, having the largest Yukawa coupling to the Higgs sector, provides a unique window into electroweak symmetry breaking and possible new physics beyond the Standard Model. Searches for rare top-quark processes are thus powerful probes of new physics. In this work, we investigate the flavor-changing neutral-current (FCNC) top-quark decays $t\to cH$ and $t\to cS$, where $S$ denotes a light scalar, at the Compact Linear Collider (CLIC) with a center-of-mass energy of $\sqrt{s}=1.5~\mathrm{TeV}$. Our analysis focuses on a kinematic regime distinct from most previous studies, in which the top quarks are typically highly boosted. To enhance signal discrimination in the boosted regime, we construct multi-channel jet images and employ a convolutional neural network (CNN) classifier to capture jet-substructure patterns relevant to the FCNC signals. Assuming an integrated luminosity of $4~\mathrm{ab}^{-1}$, we obtain the expected $95\%$ C.L. upper limit $\mathrm{BR}(t\to cH)\times \mathrm{BR}(H\to b\bar b)<5.27\times10^{-5}$. For the exotic scalar singlet, expected $95\%$ C.L. upper limits between
$3.25\times10^{-5}$ and $5.26\times10^{-5}$ are obtained for
$\mathrm{BR}(t\to cS)\times \mathrm{BR}(S\to b\bar b)$, for scalar masses
between $30$ and $80~\mathrm{GeV}$.
\end{abstract}


\maketitle
\flushbottom

\vspace{1cm}

\section{Introduction}
The top quark holds a unique position in the Standard Model (SM) due to its exceptionally large mass, which is close to the electroweak symmetry breaking scale. With a Yukawa coupling of order unity, the top quark is expected to be particularly sensitive to physics beyond the SM associated with the Higgs sector or the mechanism of mass generation.
As a result, both precision measurements of top-quark properties and searches for rare top-quark processes provide powerful probes of new physics~\cite{ParticleDataGroup:2024cfk,CMS:2024irj,ATLAS:2024dxp}.

Among the rare processes involving the top quark, flavor-changing neutral current (FCNC) decays
play a particularly important role.
In the SM, FCNC top-quark transitions are absent at tree level and are also strongly suppressed by the
Glashow--Iliopoulos--Maiani (GIM) mechanism at higher orders. 
This results in that the predicted branching ratios for FCNC decays in the SM are extremely small. For decays such as $t \to qZ$, $t \to q H$, $t \to q\gamma$ and $t \to qg$ with $q = u, c$, the values are typically in the range $10^{-17}-10^{-12}$~\cite{Eilam:1990zc, Mele:1998ag, Aguilar-Saavedra:2002lwv, Aguilar-Saavedra:2004mfd, Zhang:2013xya, Durieux:2014xla}, rendering such processes in the SM unobservable at current and foreseeable experimental facilities. However, various extensions of the SM can enhance them by several orders of magnitude~\cite{Castro:2022qkg, Durieux:2014xla, Larios:2006pb, Barros:2019wxe}.

Consequently, the observation of FCNC top-quark decays would be an unambiguous signal of physics beyond the SM (BSM), which has drawn considerable attention. Operating as a "top-quark factory", the Large Hadron Collider (LHC) has enabled a wide variety of searches for FCNC top-quark decays. Current experimental searches include top-quark decays into a light quark accompanied by a $Z$ boson~\cite{ATLAS:2012hfh,CMS:2012wao,CMS:2013knb,
ATLAS:2015vhj,CMS:2017wcz,ATLAS:2018zsq}, a photon~\cite{ATLAS:2022per},
a Higgs boson~\cite{ATLAS:2015ncl,CMS:2016obj,ATLAS:2017tas,
CMS:2017bhz,CMS:2021gfa,ATLAS:2018jqi,ATLAS:2018xxe,ATLAS:2022gzn,CMS:2024ubt},
or dilepton final states~\cite{CMS:2022ztx}, as well as broader relevant phenomenological analyses~\cite{Mahlon:1994us,Fox:2007in,Drobnak:2008br,Han:2013sea,Durieux:2014xla,Yue:2014hba,Guo:2016kea,
Chala:2018agk,Balaji:2020qjg,Cremer:2023gne,Altmannshofer:2023bfk}. In addition, extensive studies have explored top-quark decays into light
BSM particles, including charged Higgs bosons
\cite{ATLAS:2018gfm,CMS:2019bfg,CMS:2022jqc,ATLAS:2023bzb}, light (pseudo)scalars~\cite{ATLAS:2023mcc,ATLAS:2023mcc,Banerjee:2018fsx,Carmona:2022jid,Bhattacharyya:2022umc,Bahl:2023xkw,Hou:2025bjy},
fermions~\cite{Alcaide:2019pnf}, and  vector mediators
\cite{Kong:2014jwa,Kim:2014ana}. Notably, modern collider analysis methodologies leveraging machine learning further enhance sensitivity to top-quark FCNCs, a development that has garnered increasing interest~\cite{Chowdhury:2023jof,Fuks:2025qgh,Coleppa:2026fdj}.

In this work, we focus on the FCNC decays of $t\rightarrow cH$ and $t\rightarrow cS$ where $S$ denotes a light scalar singlet. The decay $t\to cH$ receives the largest enhancement in many SM extensions, with the branching ratio potentially increased by several orders of magnitude relative to the SM prediction of $~\sim 3\times 10^{-15}$. In most cases, such enhancements stem from contributions of new particles or modified couplings at the loop level. In supersymmetric scenarios (SUSY), early studies suggested $\mathrm{Br}(t \to cH) \sim 10^{-5}$~\cite{Guasch:1999jp,Cao:2007dk}; however, Higgs mass measurements and LHC constraints have reduced the MSSM prediction to
$10^{-7}~\sim 10^{-6}$~\cite{Cao:2014udj,Dedes:2014asa}, with similar values in
R-parity-violating SUSY~\cite{Eilam:2001dh}. In flavor-conserving two-Higgs-doublet models (FC 2HDM), the decay is also
loop-induced and can reach branching ratios of order $10^{-5}$ ~\cite{Atwood:1996vj}. In universal extra dimension models, the width of $t \to cH$ generally does not receive much enchancement from the contributions of Kaluza-Klein particles~\cite{Dey:2016cve}. Much larger enhancements arise in models with tree-level flavor-changing scalar couplings. In general two-Higgs-doublet models (2HDM), the branching ratio can be as large as $\mathrm{Br}(t \to cH) \sim 10^{-3}$\cite{Cheng:1987rs}, and in models with additional up-type vector-like quarks (VLQ) values of $\mathrm{Br}(t \to cH) \sim 4 \times 10^{-5}$ remain phenomenologically allowed~\cite{Aguilar-Saavedra:2006mhs}. 

On the other hand, light scalars $S$ or pseudo-scalars $a$ arise in many well-motivated extensions of the SM and may couple to top quarks, including scalar singlets in composite Higgs models, axion-like particles, which result in interesting top FCNC decays~\cite{Banerjee:2018fsx,Carmona:2022jid,Bhattacharyya:2022umc,Bahl:2023xkw,Hou:2025bjy}. In some composite Higgs models, the FCNC decay $t\to cS$ can significantly larger than those induced by the SM Higgs boson due to the relevant coupling unsuppressed by GIM mechanism~\cite{Banerjee:2018fsx,Castro:2020sba}. 
Very light (pseudo)scalar singlets associated with top FCNCs may also give rise to long-lived particle signatures~\cite{Bahl:2023xkw,Carmona:2022jid,Cheung:2024qve,Yue:2026dmy,Yue:2025wkn,Yue:2026nme}, which lie beyond the scope of this work. 

Using the full Run-2 dataset at $\sqrt{s}=13~\mathrm{TeV}$,
the ATLAS and CMS collaborations have placed stringent upper limits on
$\mathrm{Br}(t \to cH)$ at the level of $\mathcal{O}(10^{-4})$~\cite{ATLAS:2022gzn,CMS:2021gfa}.
The decays into a light scalar S and a jet are already constrained by a recent experimental search which constrains
$BR(t\to qS) \times BR(S\to b \bar{b}) \leq 10^{-4} \sim 10^{-3}$ for 20 GeV $< m_S <$160 GeV ($q=u,c$)~\cite{ATLAS:2023mcc}. These results have already provided important constraints on a variety of new physics models and highlight the need for further studies of FCNC top-quark processes at high-energy colliders.

Despite its tremendous success, the LHC environment is not ideally suited for precision studies of rare top-quark decays.
The complex hadronic environment, large QCD backgrounds, and the presence of pileup interactions limit the achievable sensitivity, particularly in fully hadronic final states. These challenges motivate the investigation of FCNC top quark decays at future lepton colliders, where a cleaner experimental environment and a well-defined initial state offer significant advantages. The Compact Linear Collider (CLIC), a proposed collider with an energy reach of up to several TeV, offers a promising platform for such investigations~\cite{Robson:2018enq, Adli:2025swq,CLICdp:2018esa}. It is designed to operate in three stages, with center-of-mass energies $\sqrt{s}$ of 380 GeV, 1.5 TeV, and 3 TeV, respectively. At the first energy stage of CLIC ($\sqrt{s}=380~\mathrm{GeV}$), considering one top decaying to $cH$ in top pair production process, the simulated study found that the sensitivity can reach
$\mathrm{Br}(t\to cH)\times \mathrm{Br}(H\to b\bar b) \lesssim 1.6\times 10^{-4}$
(at $95\%$ C.L. for $500~\mathrm{fb}^{-1}$)~\cite{CLICdp:2018esa,Zarnecki:2018wde}. In this case, the separation and assignment of the top-quark decay products remain non-trivial facing large background of SM top-pair events and severe combinatorial problem.

In this work, we investigate the discovery potential of CLIC for FCNC decays of $t\to cH $ and $t\to cS$ focusing on a distinct kinematic regime. The stage 2 of CLIC, operating at a center-of-mass energy of $\sqrt{s}=1.5~\mathrm{TeV}$ with an integrated luminosity\footnote{Original design of the integrated luminosity of CLIC at stage II is $2.5\mathrm{ab}^{-1}$~\cite{Robson:2018enq}. Benefit from improved simulation, progressed techniques and luminosity ramp-up time, the current estimation of the luminosity can reach $4~\mathrm{ab}^{-1}$~\cite{Adli:2025swq}.} of $4~\mathrm{ab}^{-1}$, yields large events samples in a qualitatively unique kinematic regime, in which top quarks and their decay products - the Higgs bosons or the light scalar $S$ - are typically highly boosted. Consequently, the decay products of a highly boosted object can be efficiently captured by a single large-$R$ jet, thus overcoming the combinatorial challenges encountered in event reconstruction for processes with high jet multiplicity events. We focus on \(e^+e^- \to t\bar{t}\) and consider the decay of one top quark via $t\to cH(S) \to c b\bar{b}$. Detailed studies of the jet substructure can thus provide a powerful handle for discriminating signal from background processes (For recent reviews of jet substructure, we refer to references~\cite{Larkoski:2017jix,Marzani:2019hun,Bonilla:2022wzp}. ). Furthermore, we enhance the analysis by employing jet image based convolutional neural networks (CNN). Jet images provide a natural representation of boosted jets~\cite{Cogan:2014oua,deOliveira:2015xxd}, while convolutional neural networks (CNNs) are ideally suited to exploit the local spatial correlations present in these images. This approach allows us to fully utilize internal information within jets and significantly enhance the sensitivity to the $t \to cH$ and $t\to cS$ decays as shown below. 

The structure of this paper is organized as follows. In the next section, we present the theoretical frameworks of top-quark FCNC interactions and summarize the current experimental constraints. In section 3, we introduce our simulation framework and provide details on the generation of both signal and background events. In Section 4, we present our analysis results for top FCNC interactions involving a Higgs boson (or a light scalar $S$) and a c quark at CLIC, with integrated luminosities of 4$\text{ab}^{-1}$ . We finally summarize our findings and conclude
in section 5.

\section{Theoretical frameworks}
Many BSM models predict significant enhancements of FCNC decays, which can arise from exotic tree-level FCNC couplings, but in most models, such enhancements stem from contributions of new particles or modified particle couplings at the loop level. 

We firstly consider the description of $tcH$ interaction and then discuss the extension with an additional scalar $S$. 
In general, Standard Model Effective Field Theory (SMEFT) ~\cite{Buchmuller:1985jz,Grzadkowski:2010es, Brivio:2017vri} provides a convenient framework to interpret potential deviations from the SM predictions in a model independent way. The SM Lagrangian ${\cal{L}}_\mathrm{SM}$ is extended by higher-dimensional operators, which are constructed from the SM fields and respect the $SU(3)_C \times SU(2)_L \times U(1)_Y$ gauge symmetry,
\begin{equation}
  \mathcal{L}_\mathrm{EFT} = \mathcal{L}_\mathrm{SM} + \sum_{D \geq 5} \frac{C_i \, \mathcal{O}_i^{(D)}}{\Lambda^{D-4}}\,.
\end{equation}
Here $\mathcal{O}_i^{(D)}$ represent gauge-invariant operators of mass dimension $D$ larger than 4, and $C_i$ are the associated dimensionless Wilson coefficients. The operators with high mass dimension are suppressed by the high energy scale of new physics. Generally, the leading new-physics effects are expected to be captured by the subset of dimension-six operators, which provides a controlled and predictive expansion valid for scales below~$\Lambda$.

In the case of top-quark FCNC interactions, several dimension-six SMEFT operators contribute to both top decay and production processes. These include operators modifying the top's interactions with the SM bosons and light quarks, as well as four-fermion interactions involving at least one top-quark field. The relevant detailed description can be found in Ref. \cite{Zhang:2014rja}.

After electroweak symmetry breaking, the SMEFT operators induce effective FCNC interactions between the top quark and the SM particles. The relevant interactions can be described in terms of physical mass eigenstates by an effective Lagrangian in which each coupling depends on one or more SMEFT Wilson coefficients, 
\begin{equation}\begin{split}
  \mathcal{L}_\text{eff} =\ & 
   - \frac{g}{2\sqrt{2}}\, \bar{q} (g^v_{qt} + g^a_{qt} \gamma_5)\, t\, h 
    - \frac{g}{2 c_W}\, \bar{t} \gamma^{\mu} (v_{tq}^{Z} - a_{tq}^{Z} \gamma_5)\, q\, Z_{\mu}\\
  & - \frac{e}{\Lambda}\, \bar{t} \sigma^{\mu\nu} (f_{tq}^{\gamma} + i h_{tq}^{\gamma} \gamma_5)\, q\, A_{\mu\nu}
    - \frac{g}{2 c_W \Lambda}\, \bar{t} \sigma^{\mu\nu} (f_{tq}^{Z} + i h_{tq}^{Z} \gamma_5)\, q\, Z_{\mu\nu} \\
  & - \frac{g_s}{\Lambda}\, \bar{t} \sigma^{\mu\nu} T^A (f_{tq}^{g} + i h_{tq}^{g} \gamma_5)\, q\, G_{\mu\nu}^A 
    + \text{H.c.}
\end{split}\label{effLag}\end{equation}

The effective couplings $g_{qt}^{v,a}$, $v_{tq}^{Z}$, $a_{tq}^{Z}$, $f_{tq}^{V}$ and $h_{tq}^{V}$ (with $V = \gamma, Z, g$) parametrize the strength of the FCNC interactions of the top quark with the Higgs boson $h$, the $Z$ boson, the photon $A$ and the gluon $g$, respectively. And the corresponding gauge boson field-strength tensors are represented by $A_{\mu\nu}$, $Z_{\mu\nu}$ and $G_{\mu\nu}^A$. Here $q= u,c$ denotes a light up-type quark. Although the effective Lagrangian does not reflect the full gauge structure of SMEFT at high energies, it serves as  a convenient and widely adopted framework for describing FCNC processes in the top-quark sector~\cite{Aguilar-Saavedra:2004mfd, Aguilar-Saavedra:2008nuh, Durieux:2014xla,Bhattacharya:2023beo}. In the following subsection 4.1, we only retain the $\bar{c}th$ terms in the effective Lagrangian and neglect all other terms, focusing on the FCNC decay $t\to cH$ with following decay of $H\to b\bar{b} $.

Similar to the above top FCNC effective Lagrangian, the relevant interactions can be readily described for an extended scalar $S$. The effective Lagrangian of $tcS$ interaction can be constructed directly in the broken phase, as follows~\cite{Franceschini:2016gxv,Banerjee:2018fsx,Castro:2020sba}:
\begin{equation}\begin{split}
  \mathcal{L}_{S} =\ &
  - \frac{1}{\sqrt{2}}\, S\, \bar{q}\left(y^{S}_{qt} + i\, y^{P}_{qt}\gamma_{5}\right)t \\
  & - \frac{1}{\sqrt{2}}\, S\, \bar{b}\left(y^{S}_{bb} + i\, y^{P}_{bb}\gamma_{5}\right)b
  + \text{H.c.},
\end{split}\end{equation}
where $q=u, c$ denotes a light up-type quark. Here, the effective couplings $y^{S}_{qt}$ and $y^{P}_{qt}$ parametrize the scalar and pseudoscalar flavor-changing interactions of the top quark with the new scalar $S$, while $y^{S}_{bb}$ and $y^{P}_{bb}$ describe the flavor-conserving couplings of $S$ to bottom quarks. In another branch of this work, we consider the case in which $m_S<m_t-m_c$, allowing the rare decay channel $t\to cS$ to open. We focus on the decay mode $t\to cS$ with $S\to b\bar b$, which leads to a final-state topology similar to that of $t\to cH\to c b\bar b$, although there exist some kinematic differences. 

\section{Event simulation and jet-image based CNN network}

\begin{figure}[htb]
  \centering
  \includegraphics[width=0.60\textwidth]{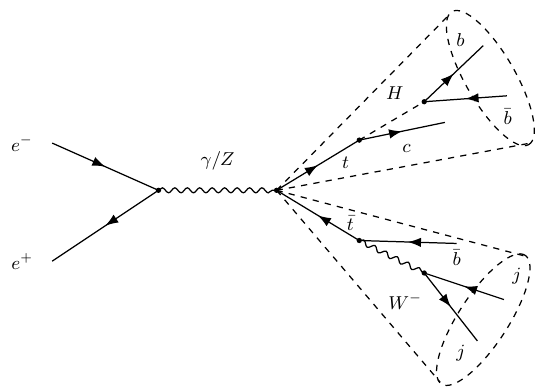}
  \caption{Representative Feynman diagrams for the signal. Here, the large cones with dash lines indicate that we study the highly boosted top quarks and aim to capture them using a large jet radius.}
  \label{fig:feyn}
\end{figure}

The analysis is performed at CLIC with a center-of-mass energy of $\sqrt{s} = 1.5$~TeV.
The signal process considered is the production of a top--antitop pair, in which one of the top quarks undergoes the FCNC decay $t \to cH$ or $t \to cS$, followed by the hadronic decay $H(S) \to b\bar{b}$. The dominant Standard Model background arises from top-quark pair production with the hadronic decay $t \to Wb \to jjb$. Representative Feynman diagram for the signal is shown in Fig.~1. Notably, at CLIC with $\sqrt{s}=1.5 \text{TeV}$ the relevant processes reside in a distinct kinematic regime different from those in most studies. In this regime, top quarks are typically highly boosted, and their decay products can be identified as fat jets. In our simulation, the original parton level event is generated by
\texttt{MadGraph5\_aMC@NLO}~v3.3.0~\cite{Alwall:2014hca}.
and then interfaced to
\texttt{Pythia}~8.313~\cite{Bierlich:2022pfr} for parton showering and hadronization. Final-state particles after hadronization with HepMC format ~\cite{Dobbs:2001ck} are passed to Fastjet~\cite{Cacciari:2011ma} for jet reconstruction.
For both the signal and background processes, samples of $10^6$ events are generated.

\subsection{Parton level analysis}

\textbf{ Kinematic distributions for $t\to cH$}
\\
\begin{figure}[htb]
    \centering
    \includegraphics[width=0.38\textwidth]{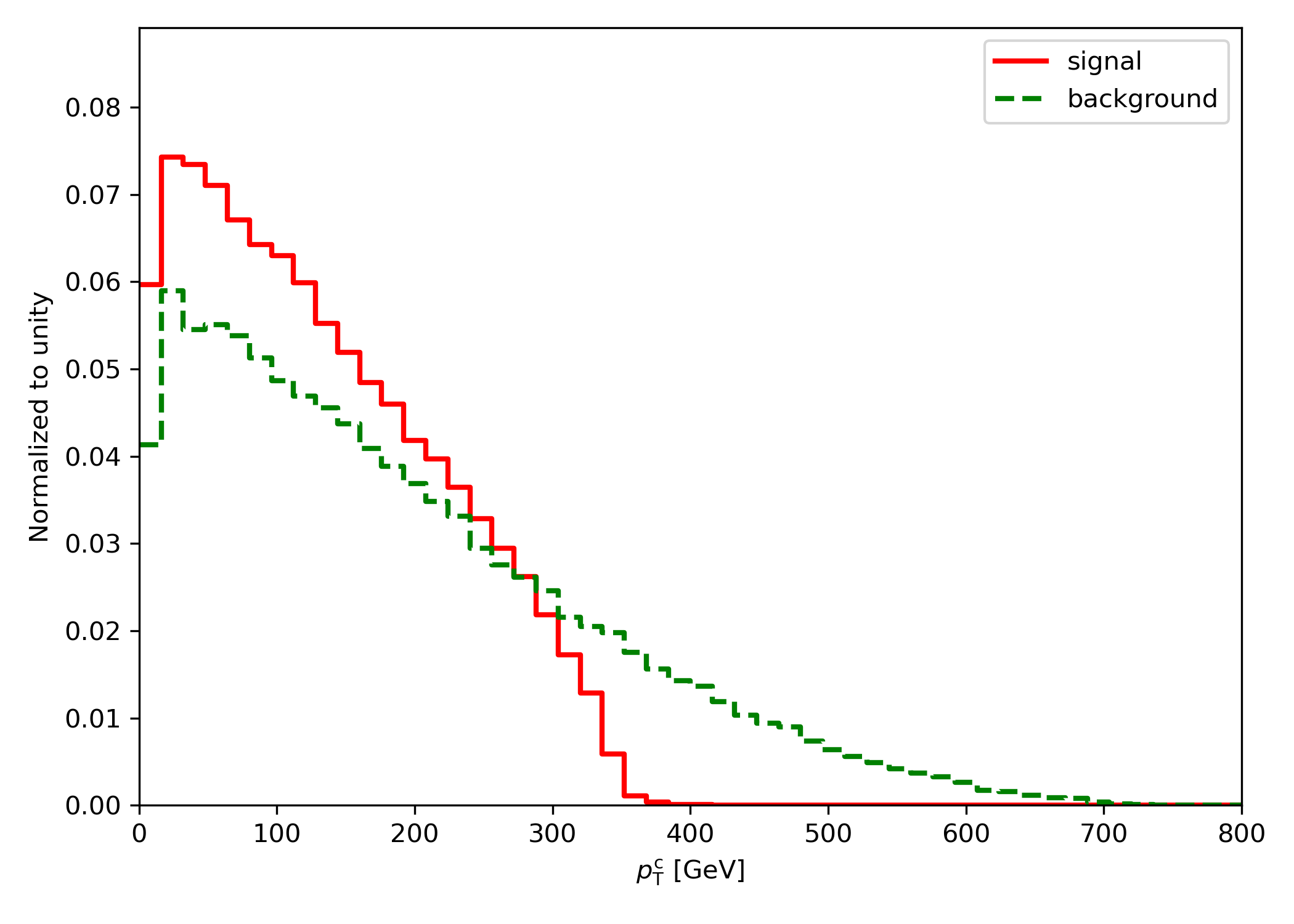}
    \hspace{0.02\textwidth}
    \includegraphics[width=0.38\textwidth]{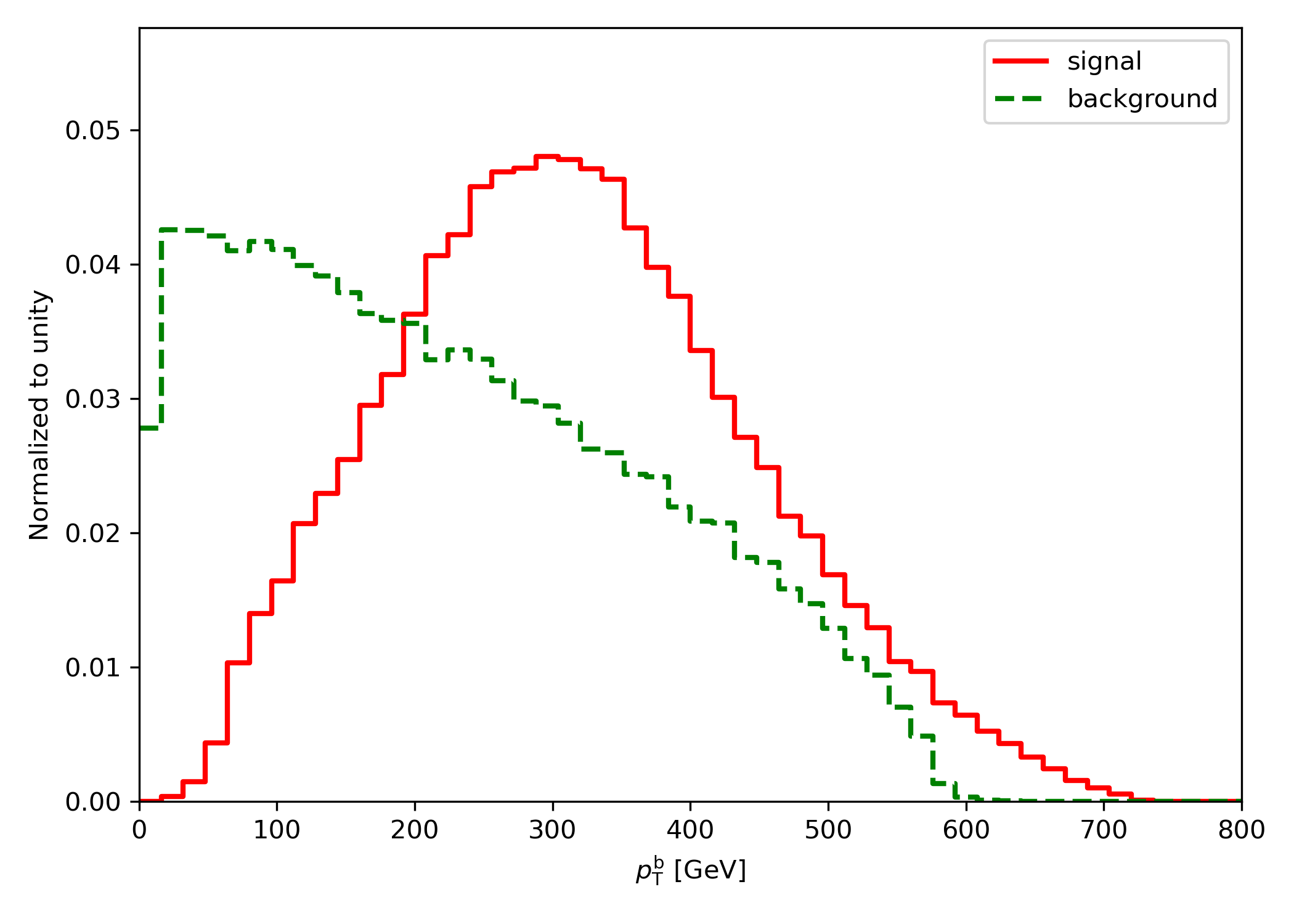}
    \vspace{0.02\textwidth}
    \includegraphics[width=0.38\textwidth]{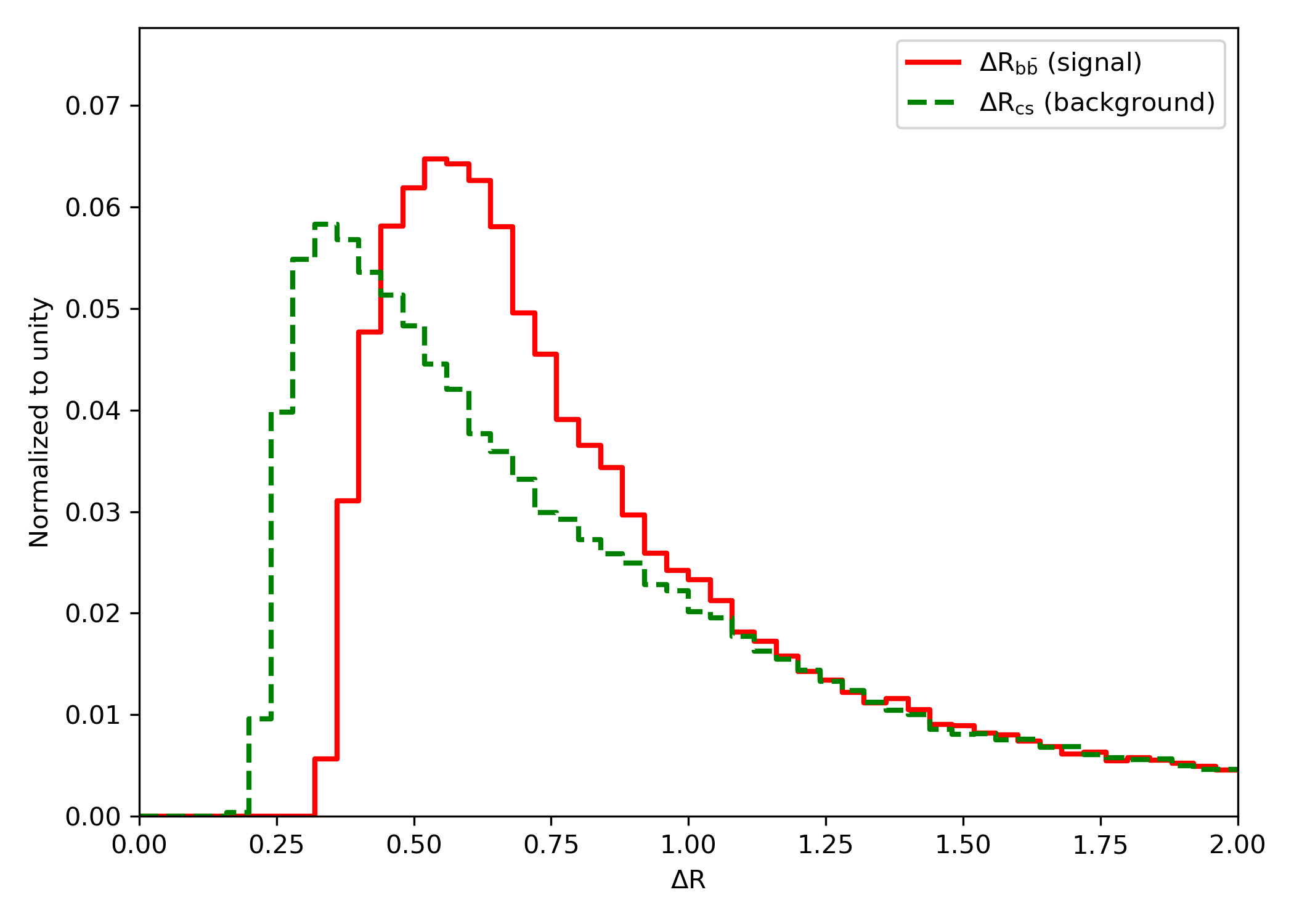}
    \hspace{0.02\textwidth}
    \includegraphics[width=0.38\textwidth]{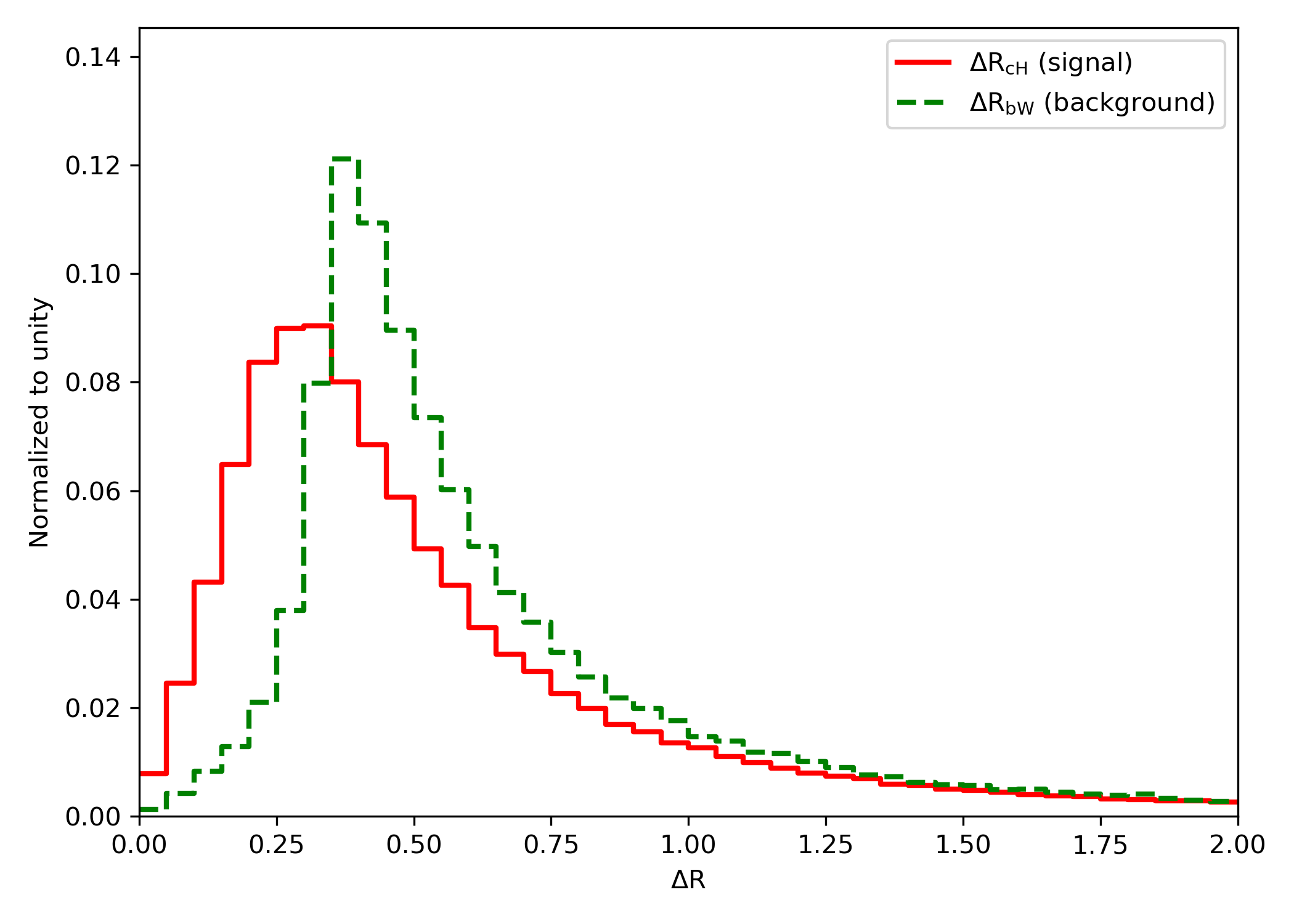}
    \hfill
    
    \caption{Normalized kinematic distributions for the signal and background. Top-left: $p_T$ distribution of the $c$-quark originates from the top decay in the signal and from the $W$-boson decay in the background. Top-right: $p_T$ distribution of the $b$-quark originates from the Higgs-boson decay in the signal and from the background.  Bottom-left: The angular separation $\Delta R(b,\bar b)$ in the signal and $\Delta R(c,\bar s)$ in the background. Bottom-right: $\Delta R(c,H)$ in the signal and $\Delta R(b,W)$ in the background.}
    \label{fig:parton_pt}
\end{figure}

\begin{figure}[!htbp]
    \centering
    \includegraphics[width=0.65\textwidth]{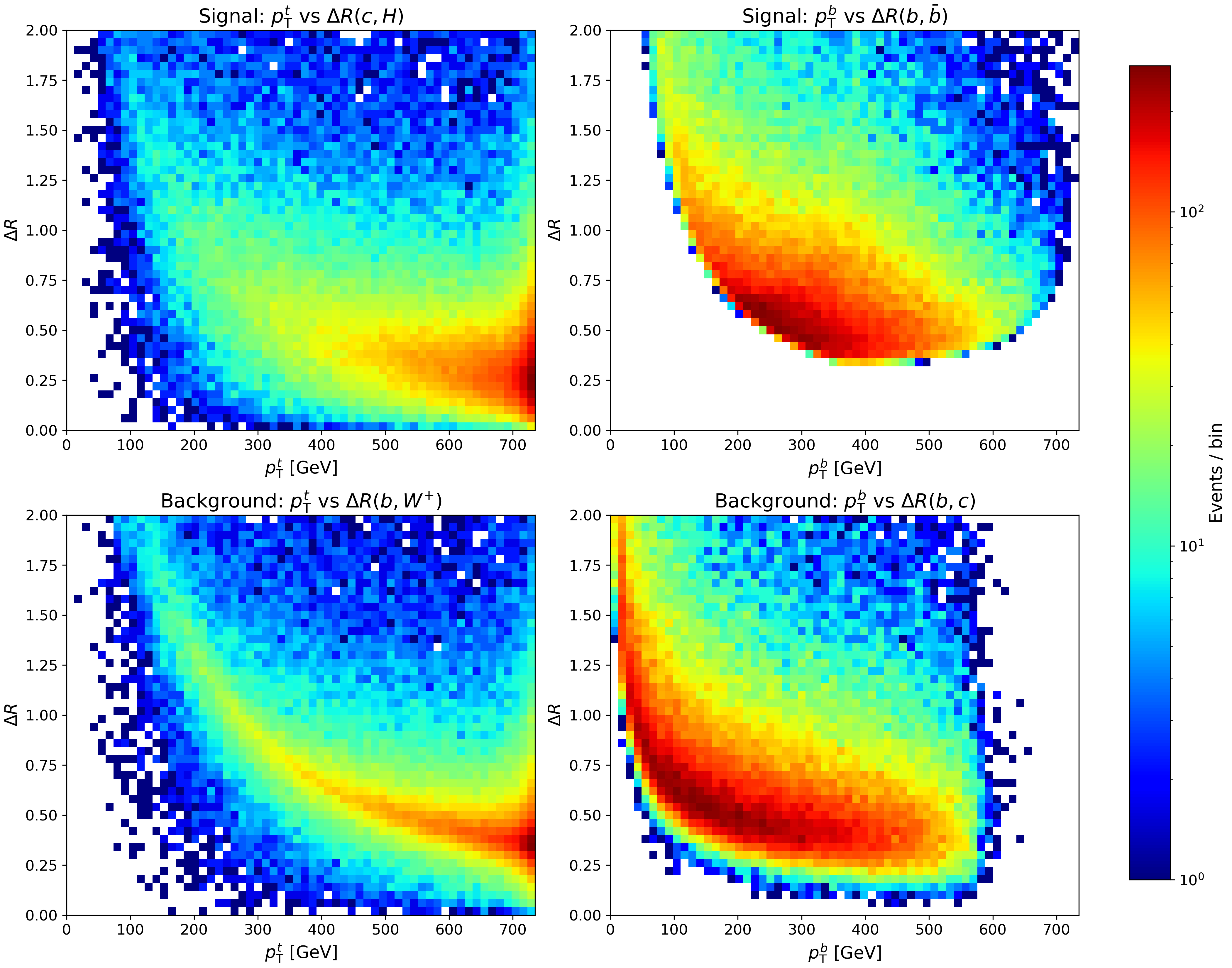}
    \caption{
    Distributions in the $p_T$--$\Delta R$ plane for the $t\to cH$ signal and the $t\to bW^+$ background events. 
    Top-left: $p_T^t$ versus $\Delta R(c,H)$ for the signal. 
    Top-right: $p_T^b$ of the leading $b$-quark from $H\to b\bar b$ versus $\Delta R(b,\bar b)$ for the signal. 
    Bottom-left: $p_T^t$  versus $\Delta R(b,W^+)$ for the background. 
    Bottom-right: $p_T^b$ of the $b$-quark from the top decay versus $\Delta R(c,b)$ for the background.
}
    \label{fig:parton_ptdr}
\end{figure}
We firstly examine several parton level kinematic observables for the signal and background processes. As shown in the Fig.~\ref{fig:parton_pt}, the features of kinematic distributions for signal and background are distinctly different. The charm quark in the signal peaks at low $p_T$, while that in the background exhibits a harder distribution with a longer  tail. The high-$p_T$ tail of the charm quark comes from the cumulative boost of the cascade decay. The maximum lab-frame $p_T$ occurs when all boosts are nearly aligned: A highly boosted top quark decays to a $W$ near its flight direction, and that $W$ subsequently decays to a $c$ quark along the $W$'s flight direction. 
In addition, the $b$-quark spectrum shows a pronounced difference. The $p_T$ distribution of the harder $b$-quark for signal exhibits a relatively broad peak around 200-300 GeV and a long tail due to the boosted effects, while background events tend to concentrate in low $p_T^b$ region.

We also study the angular separations between the relevant decay products. As shown in Fig.~\ref{fig:parton_pt}, the distributions of $\Delta R(b,\bar b)$ in the signal and $\Delta R(c,\bar s)$ in the background are both concentrate in the small-$\Delta R$ region due to the boost effect. The separation $\Delta R(c,\bar s)$ peaks at a smaller value than $\Delta R(b,\bar b)$ because the corresponding mother particle, the $W$ boson, has a smaller mass $m_W$ and is therefore more effectively boosted. In addition, a clearer separation is observed for the primary decay products: the signal distribution of $\Delta R(c,H)$ peaks at a smaller value than the background distribution of $\Delta R(b,W)$. Because the $c$ and $b$ quarks are very light compared to the top quark mass, the angular separation mainly depends on the mass of the heavy particles, i.e. the Higgs boson and the W boson. In the top quark rest frame, the lighter $W$ has a larger velocity than the Higgs. After a strong boost to the laboratory frame, this larger velocity gives the $W$ a greater “resistance” to being aligned, leaving it less collimated with its partner quark than the Higgs is with its own.

The kinematic features are further illustrated in the two-dimensional $p_T$--$\Delta R$ distributions shown in Fig.~\ref{fig:parton_ptdr}. The two left panels demonstrate the event density distribution in the plane of top's transverse momentum $p^t_T$ and the angular separation $\Delta R$ between its decay products. Most signal events are concentrated in the region with $p_T\gtrsim 400~\mathrm{GeV}$ and $\Delta R\lesssim 1$, indicating that the decay products are typically contained within a compact angular region. 
This observation motivates the boosted-top selection and the use of large-$R$ jet reconstruction in the following analysis.
For background events, highly boosted top quarks yield similar angular separations, which poses a challenge to our analysis.

The two sub-figures in the right column of Fig.~\ref{fig:parton_ptdr} show the correlation between the leading $b$-quark transverse momentum and the corresponding angular separation. 
For the signal, the leading $b$-quark from $H\to b\bar b$ is mainly populated in the range $p_T^b\simeq 100$--$400~\mathrm{GeV}$, and the $b\bar b$ pair is concentrated around $\Delta R(b,\bar b)\simeq 0.3$--$0.75$. 
By contrast, the background exhibits a broader $p_T^b$ distribution, together with a wider $\Delta R(c,b)$ spread. 
This difference reflects the distinct decay topology of the background and implies a less localized jet substructure pattern compared with the signal. 

Based on above kinematic features from parton-level analysis, we will employ a large jet radius $R=1.0$ and select fat jet with $p_T\geq 400$ GeV in the following analysis. It is expected that refined jet-image learning can better separate signal events from the backgrounds.

\vspace{0.5cm}
\textbf{ Kinematic distributions for $t\to cS$}
\\

\begin{figure}[!htbp]
    \centering
    \includegraphics[width=0.38\textwidth]{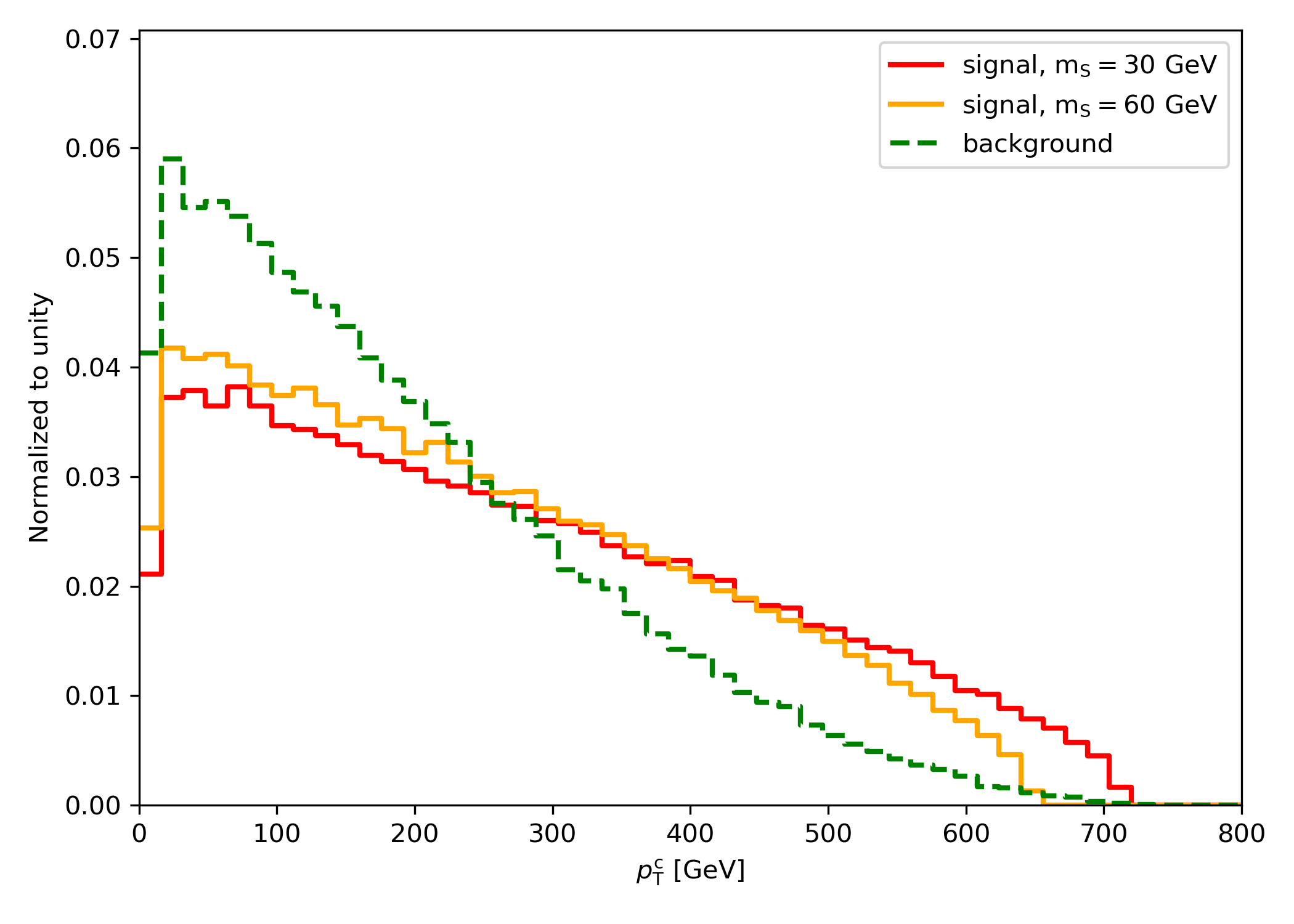}
    \hspace{0.02\textwidth}
    \includegraphics[width=0.38\textwidth]{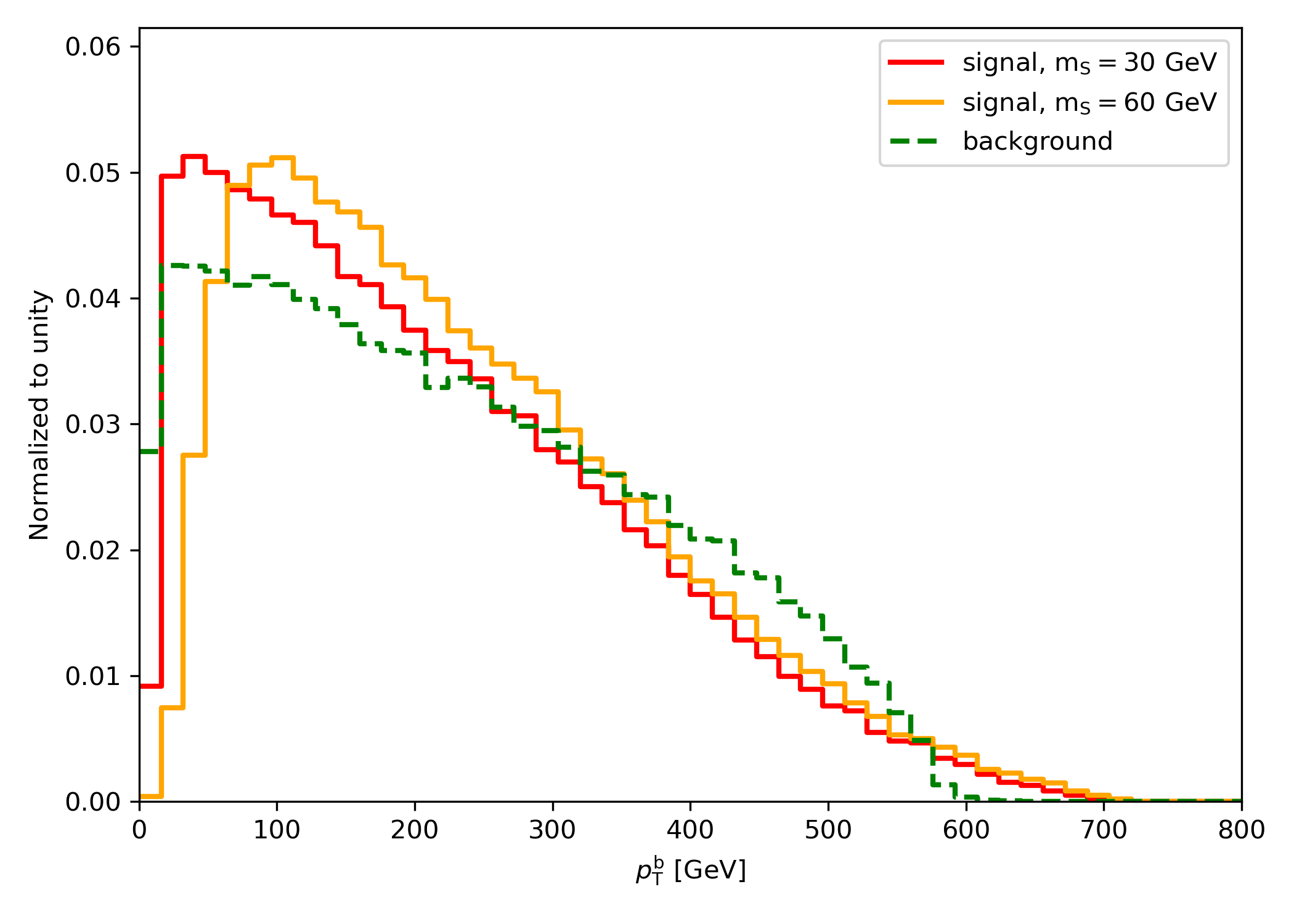}

    \vspace{0.02\textwidth}

    \includegraphics[width=0.38\textwidth]{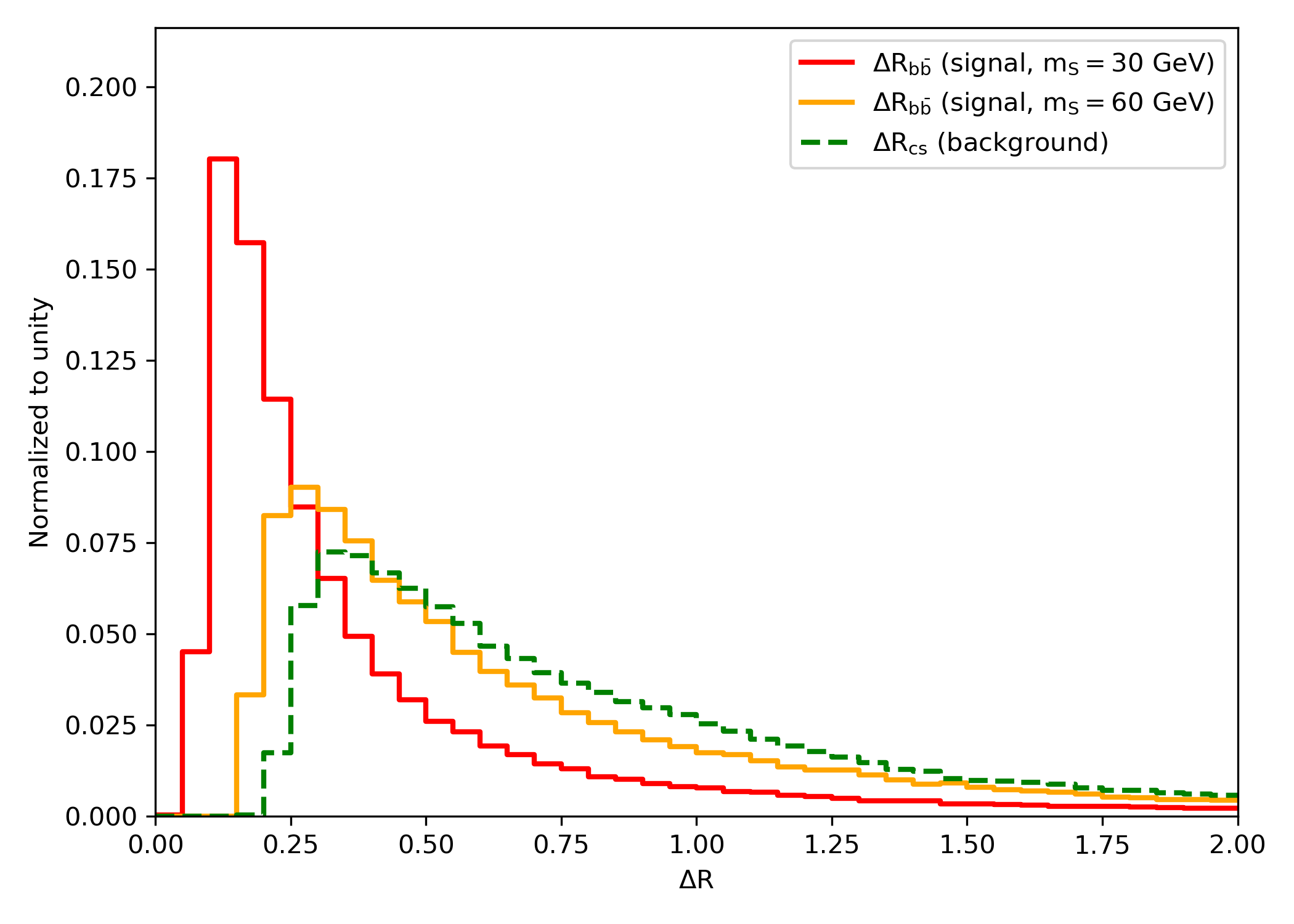}
    \hspace{0.02\textwidth}
    \includegraphics[width=0.38\textwidth]{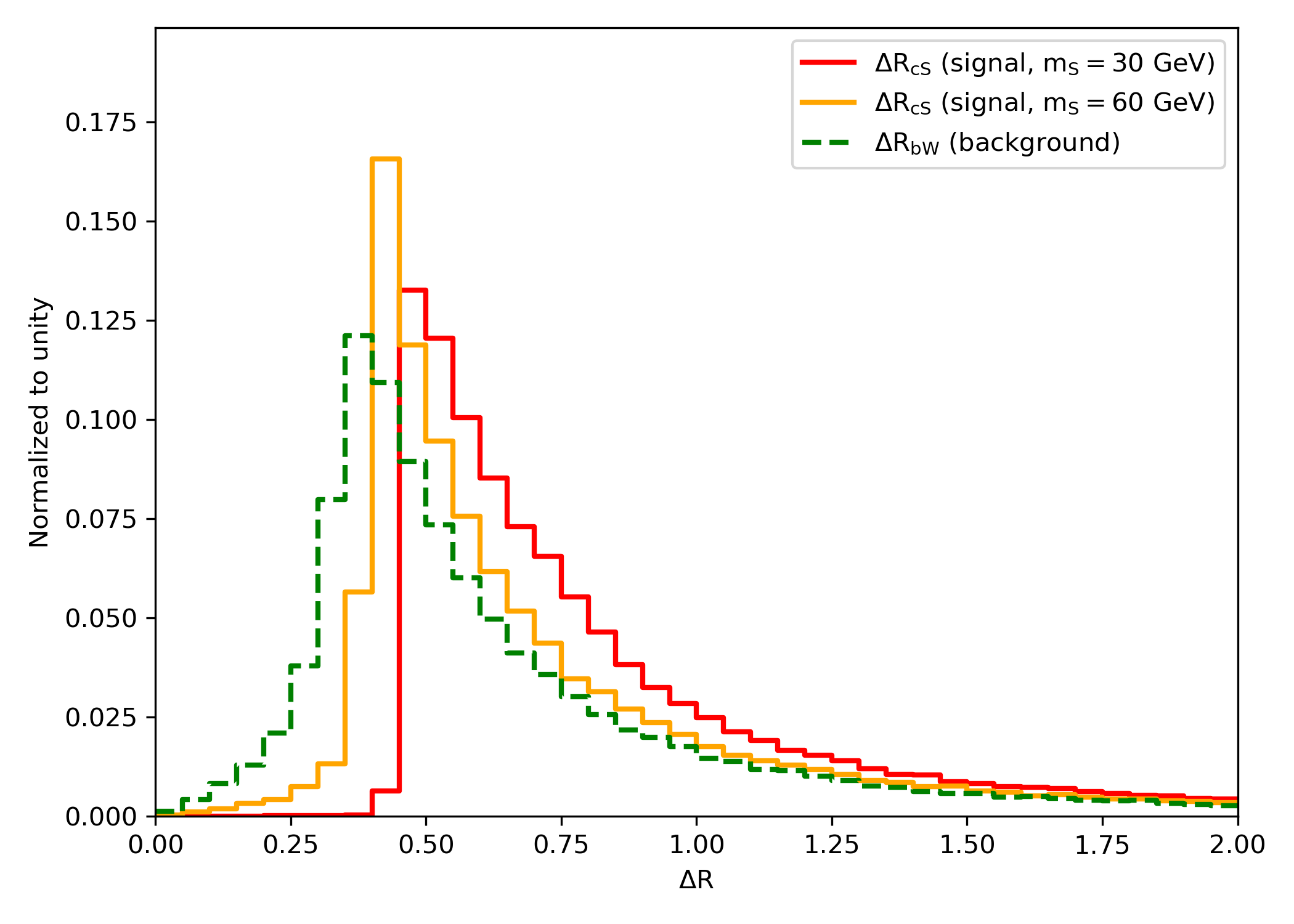}
    \caption{
        Normalized parton-level kinematic distributions for the
        $t\to cS$ signal and the background.
        The blue, green, and red curves denote the
        $\mathrm{m}_{\mathrm{S}}=30~\mathrm{GeV}$ signal,
        $\mathrm{m}_{\mathrm{S}}=60~\mathrm{GeV}$ signal, and background,
        respectively.
        Top-left: $p_{\mathrm{T}}^{\mathrm{c}}$.
        Top-right: $p_{\mathrm{T}}^{\mathrm{b}}$ for the leading $b$-quark.
        Bottom-left: $\Delta \mathrm{R}(b,\bar b)$ for the signal and
        $\Delta \mathrm{R}(c,\bar s)$ for the background.
        Bottom-right: $\Delta \mathrm{R}(c,S)$ for the signal and
        $\Delta \mathrm{R}(b,W)$ for the background.
    }
    \label{fig:parton_kin_tcs}
\end{figure}

As for the $t\to cS$ FCNC process, the kinematic distributions show certain distinctive features.  
Treating the scalar mass $m_S$ as a free parameter, we consider the mass range $30 ~\rm{GeV} <m_S<80 ~\rm{GeV}$ and present kinematic distributions at two benchmark points, $m_S=30$ GeV and $m_S=60$ GeV for illustration in this section. 

\begin{figure}[!htbp]
    \centering
    \includegraphics[width=0.62\textwidth]{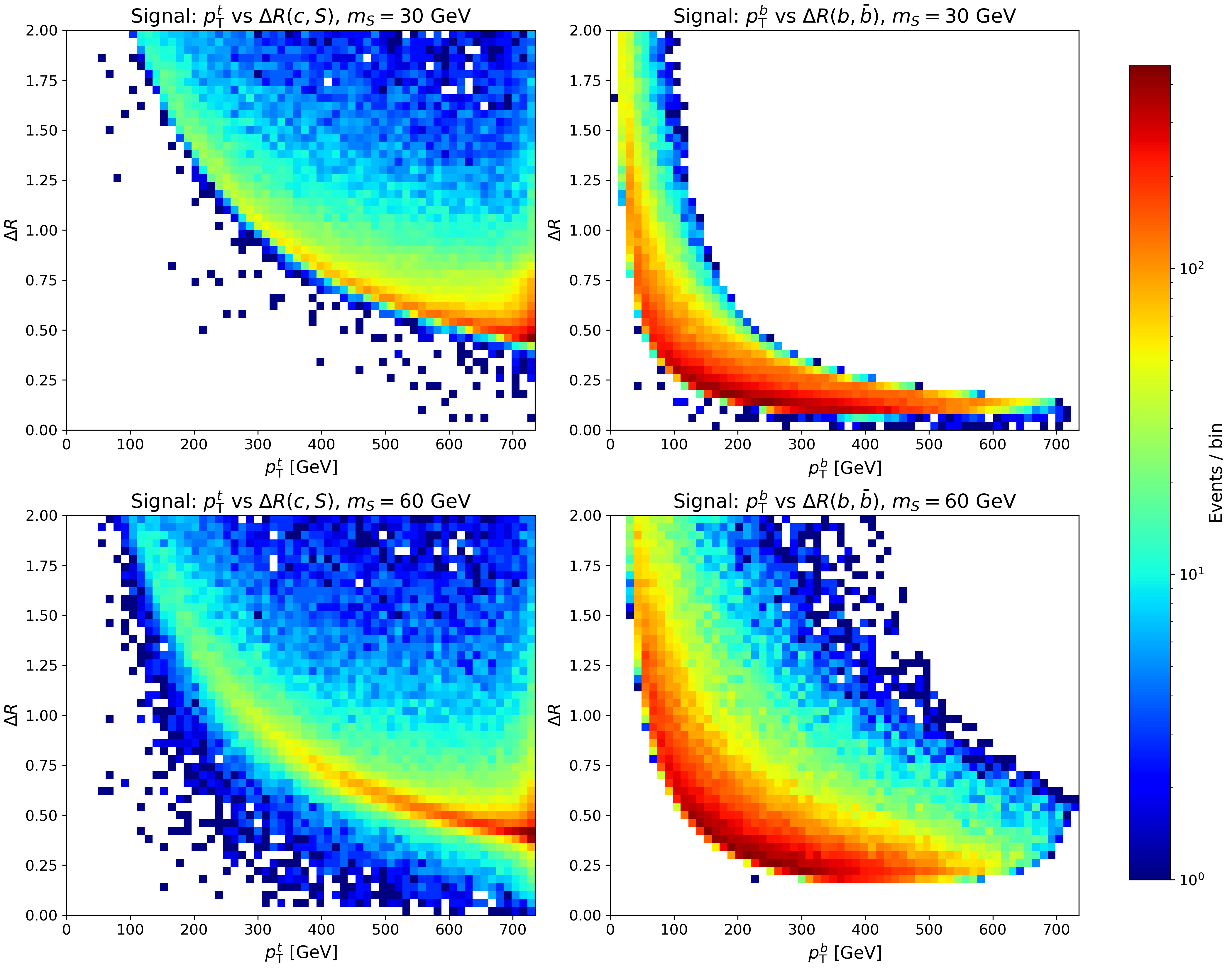}
    \caption{Distributions in the $p_T$--$\Delta R$ plane for the $t\to cS$, $S\to b\bar b$ signal process at two benchmark scalar masses. 
    Top-left: $p_T^t$ versus $\Delta R(c,S)$ for $m_S=30~\mathrm{GeV}$. 
    Top-right: $p_T^b$ of the leading $b$-quark from $S\to b\bar b$ versus $\Delta R(b,\bar b)$ for $m_S=30~\mathrm{GeV}$. 
    Bottom-left: $p_T^t$ versus $\Delta R(c,S)$ for $m_S=60~\mathrm{GeV}$. 
    Bottom-right: $p_T^b$ of the leading $b$-quark from $S\to b\bar b$ versus $\Delta R(b,\bar b)$ for $m_S=60~\mathrm{GeV}$.
    }
    \label{fig:parton_ptdr_tcs}
\end{figure}

We examine the transverse-momentum and angular distributions of the partons, shown in Fig.~\ref{fig:parton_kin_tcs}. 
A clear scalar-mass dependence is observed in these observables. 
As $m_S$ increases, the $p_T$ distribution of the charm quark shifts toward lower values, while the $p_T$ distribution of the leading $b$-quark tends to become harder. 
For a new scalar with a mass $m_S$ in the range we considered, its decay products tend to exhibit a smaller $\Delta R$ compared to those of the Higgs boson, as light particles are more easily boosted. 
For the benchmark point $m_S=30$ GeV, the two $b$-quarks from the scalar decay are highly collimated, so that the distribution of $\Delta R(b,\bar b)$ peaks around $\Delta R\sim 0.12$. 
For a larger mass, $m_S=60$ GeV, the two $b$-quarks become less collimated and the $\Delta R(b,\bar b)$ distribution shifts toward larger values. As for the separation between top's decay products, the distribution of $\Delta R(c,S)$ tends to move toward larger values as $m_S$ decreases. Since the light quarks are ultra-relativistic, the separation is controlled by the heavy particle. In the top quark rest frame, the lighter $S$ has a larger velocity than the heavier $S$ and $W$ cases. After a strong boost from top's high $p_T$, the lighter $S$ boson retains more of its rest-frame velocity, leading to a wider separation and resisting perfect alignment.

The two-dimensional distributions shown in Fig.~\ref{fig:parton_ptdr_tcs} provide additional insights. The general boosted-decay behavior remains visible: events with larger transverse momentum tend to have smaller angular separations. And the mass dependence is further illustrated by the two-dimensional $p_T$--$\Delta R$ distributions. 
As shown in the figure, the $m_S=30$ GeV events are compressed into a narrow band in the small-$\Delta R$ regions. In particular, for the $b\bar b$ system, the region with simultaneously large $p_T$ and large $\Delta R(b,\bar b)$ is almost empty, which is a typical feature of a highly boosted light resonance. For $m_S=60$ GeV, the event population becomes more broadly distributed in $\Delta R(b,\bar b)$, indicating a less boosted effect.    

Compared with the background distributions in Fig.\ref{fig:parton_ptdr}, these kinematic features reveal the differences between low-$m_S$ signal events and background events, further motivating our subsequent jet-image analysis. In the subsequent jet-image learning targeting the $t\to cS$ signal, we will also adopt pre-selection cuts with $R=1.0$ and $p_T\geq400$ GeV for fat jets.

\subsection{Jet reconstruction and jet image construction}
Jets are clustered using the \texttt{FastJet} package~\cite{Cacciari:2011ma}, employing sequential recombination algorithm of Valencia~\cite{Boronat:2016tgd} with an emphasis on large-R jets, which are particularly suited for capturing the collimated decay products of boosted objects in our study. 
Here, the jet radius $R=1.0$ is used for the jet clustering. 
The leading large-$R$ jet is selected as the signal candidate and is subsequently used for jet image construction.
We require the transverse momentum of jets to satisfy
\(p_T > 400~\mathrm{GeV}\) and its pseudorapidity to be within \(|\eta| < 2.0\),
ensuring good containment of the jet constituents within the detector acceptance~\cite{deOliveira:2015xxd,Kasieczka:2017nvn}. These requirements correspond to the boosted kinematic regime  at
\(\sqrt{s} = 1.5\)~TeV and help suppress all background contributions except top pair events.
Furthermore, the jet mass is required to lie within a window
\(140~\mathrm{GeV} < m_J < 200~\mathrm{GeV}\).
Jets satisfying these criteria are identified as signal candidates and are
used as inputs for the jet image construction and the subsequent
machine learning based classification.
Previous studies have shown that boosted jets originating from
$t\to cH$ and $t\to bW$ decays exhibit very similar three-prong substructure
patterns~\cite{Chowdhury:2023jof}, making their discrimination particularly
challenging when relying solely on low-level kinematic information. Motivated by this observation, we augment the jet image representation by
including flavor-sensitive information in additional image channels.

Each selected fat jet is represented as a multi-channel jet image by projecting its constituents onto a regular $45\times45$ grid in the $\eta$--$\phi$ plane. The image covers the region $\eta,\phi\in[-1.125,1.125]$ relative to the jet axis, corresponding to a pixel granularity of $\Delta\eta\times\Delta\phi=0.05\times0.05$. The grid is centered on the jet axis.

Following standard jet image preprocessing procedures, the image is translated to center the pixel with the largest transverse momentum, and rotated such that the leading substructure feature is aligned directly below the leading $p_T$ core of the jet image in the $\eta$--$\phi$ plane~\cite{deOliveira:2015xxd,Kasieczka:2017nvn}.

Based on this common geometric framework, three complementary jet image channels are constructed. The first channel encodes the transverse-momentum flow, with each pixel intensity defined as the scalar sum of the transverse momenta of the constituents in that pixel and normalized to unit maximum. The second channel incorporates displaced-track information, where the pixel
intensity is given by the maximum absolute transverse impact parameter
$|d_0|$ of charged particles within the pixel, with $|d_0|$ capped at
$d_0^{\max}=1500~\mathrm{mm}$~\cite{CLICdp:2018vnx}. The third channel is a spatial $b$-tagging score map. Heavy-flavour information associated with the selected jet is mapped onto the same $\eta$--$\phi$ grid, and localized responses are deposited around the corresponding pixel positions. The resulting $b$-score map is normalized to unit maximum.

All three channels share identical coordinate systems and pixelization schemes, allowing correlated spatial patterns of kinematic and flavor-sensitive information to be represented in a unified image format. These multi-channel jet images are used as the inputs to the CNN classifier described below.

\subsection{CNN architecture and training setup}

To analyze the jet images constructed above, we employ a convolutional neural network (CNN) classifier~\cite{LeCun:2015pmr,Pedregosa:2011ork}. The three physics-motivated image channels, namely the transverse-momentum flow channel, the displaced-track $d_0$ channel, and the spatial $b$-score channel, are further augmented by two auxiliary coordinate channels encoding the normalized pixel positions in the $\eta$ and $\phi$ directions. The coordinate channels are defined on a fixed regular grid spanning $[-1,1]$ in both directions. The final input to the CNN is therefore a $45\times45\times5$ tensor.

The network is a residual convolutional architecture with squeeze-and-excitation (SE) blocks\cite{Hu:2017jsr}. The input is processed by a $3\times3$ convolution (32 channels), followed by layer normalization and LeakyReLU activations with $\alpha=0.2$. The backbone consists of four stages with channel dimensions 32, 64, 128, and 256, each containing two residual blocks. Each block consists of two $3\times3$ convolutions with layer normalization and LeakyReLU, followed by an SE block.  When the spatial resolution or channel dimension changes, the shortcut branch uses a $1\times1$ projection with matching stride.

Downsampling is applied at the beginning of stages 2--4 via stride-2 convolutions, reducing the spatial resolution from $45\times45$ to $23\times23$, $12\times12$, and $6\times6$. Dropout is applied after residual addition with rates 0.03, 0.06, 0.10, and 0.15 for stages 1--4. L2 regularization is applied to all convolutional layers with coefficients $10^{-4}$, $2\times10^{-4}$, $4\times10^{-4}$, and $8\times10^{-4}$ for stages 1--4, respectively.

The final feature maps are aggregated by global average pooling and passed to a fully connected layer with 128 units, followed by layer normalization, LeakyReLU ($\alpha=0.2$), and dropout (0.25). L2 regularization with coefficient $10^{-3}$ is applied to this dense layer. A sigmoid output node produces the binary classification score.

Input preprocessing is applied channel-wise. The $p_T$ and $d_0$ channels are non-negative by construction and are transformed using $\log(1+x)$ to reduce the dynamic range. Any residual negative entries, which may arise from intermediate numerical handling or signed-convention inputs, are set to zero before the transformation. The $b$-score channel is clipped to be non-negative without logarithmic transformation. The coordinate channels are fixed for all samples. During training, only the training samples are augmented by a left--right flip applied independently to each sample with probability 0.5; all channels are flipped consistently and the horizontal coordinate channel is multiplied by $-1$.

The dataset used for training are randomly split into training and validation sets with fractions 70\% and 30\%, respectively, using a fixed seed of 2026. Training samples are reshuffled at each epoch. Class weights are computed from the training set and applied during optimization. An independent
test sample is kept separate from the training and validation procedure and is used only for the final performance evaluation.

Training is performed for up to 300 epochs with a global batch size of 8192. The loss is binary cross-entropy with label smoothing $\epsilon=0.02$, implemented within the loss by shifting targets toward 0.5. Optimization uses AdamW with $\beta_1=0.9$, $\beta_2=0.999$, and weight decay $10^{-4}$; both L2 kernel regularization and AdamW weight decay are applied. The learning rate
follows cosine decay from $3\times10^{-4}$ to $1.5\times10^{-5}$ over the full training. Model performance is monitored using accuracy and AUC, and early stopping monitors the validation AUC with a patience of 80 epochs. After training, the classifier threshold used for the final sensitivity estimate is chosen separately according to the expected exclusion performance, rather than being fixed by the epoch with the highest validation AUC.

\begin{figure*}[!htbp]
  \centering
  \includegraphics[
    width=0.9\textwidth,
    trim=0 60 0 0,
    clip
  ]{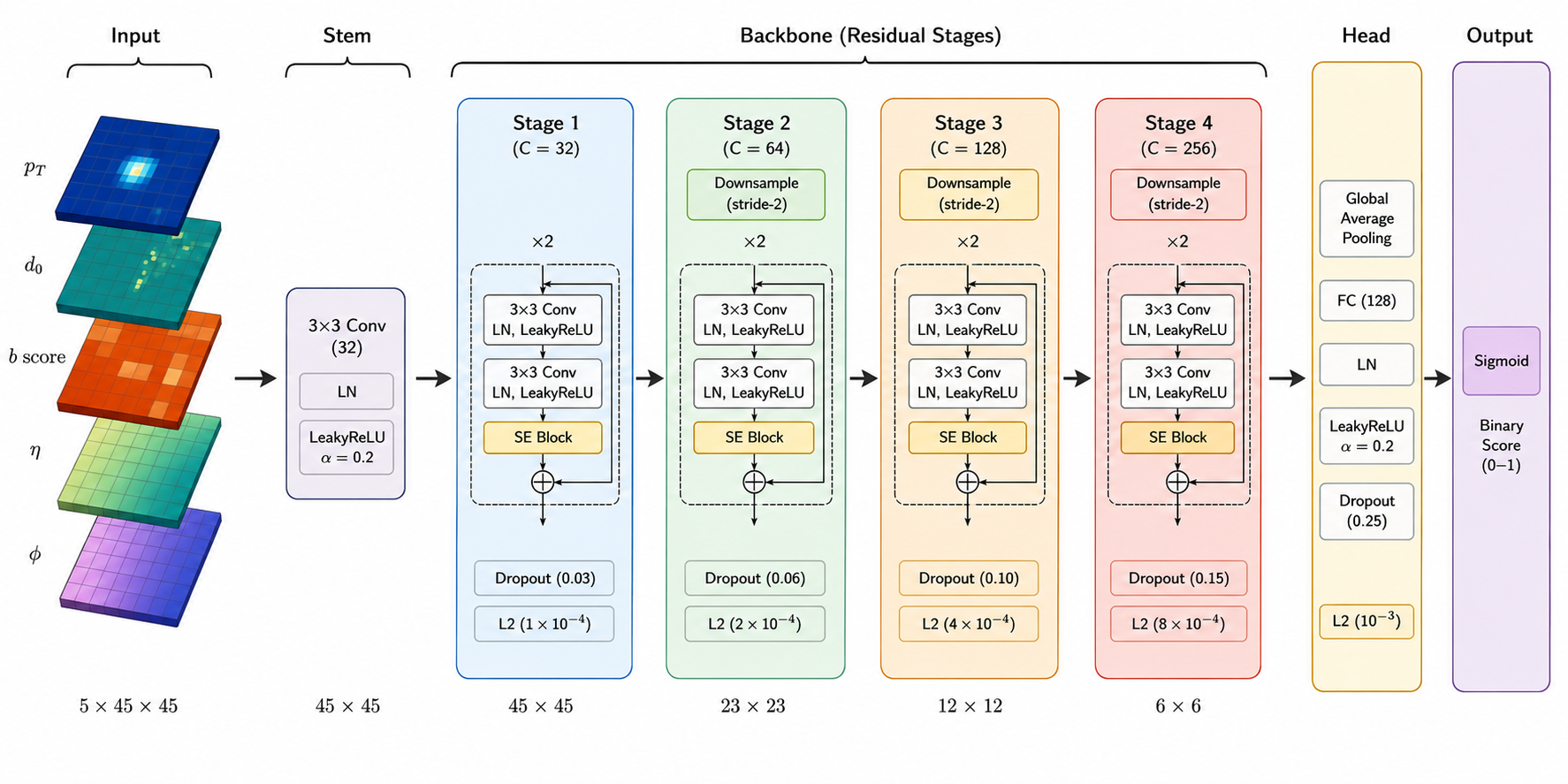}
  \caption{
  Schematic illustration of the CNN architecture used for the multi-channel
  jet-image classification. The input jet image is represented as a
  $45\times45\times5$ tensor. The network consists of
  an initial $3\times3$ convolution, followed by four residual stages with SE
  blocks and channel dimensions 32, 64, 128, and 256. The final feature maps
  are aggregated by global average pooling and passed through a dense layer to
  produce the binary classification score.
  }
  \label{fig:Architecture}
\end{figure*}

The overall network architecture is summarized in Fig.~\ref{fig:Architecture}.

\section{Collider analysis}
In this section, we present the results obtained from our analysis after training the data with our network. The studies for $t\to ch$ and $t\to cS$ are presented in two subsections, respectively.

\subsection{$t\to cH$ analysis}
\begin{figure*}[!htbp]
  \centering
  \begin{subfigure}{0.30\textwidth}
    \centering
    \includegraphics[width=\linewidth]{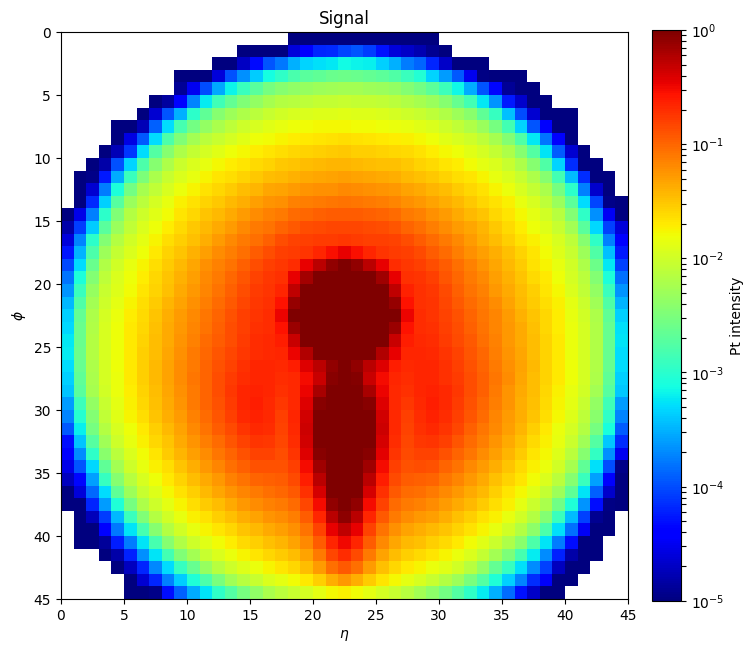}
    \caption{Signal, $p_T$}
  \end{subfigure}
  \hspace{0.015\textwidth}
  \begin{subfigure}{0.30\textwidth}
    \centering
    \includegraphics[width=\linewidth]{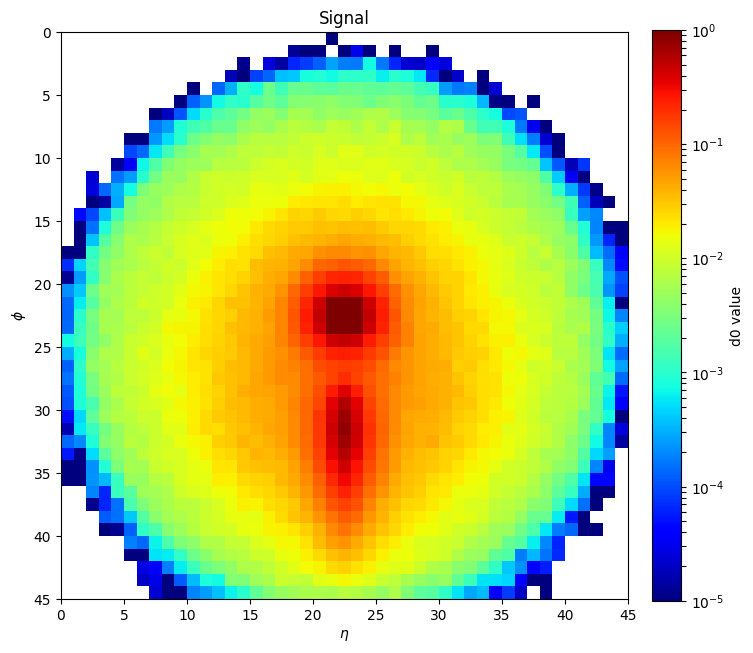}
    \caption{Signal, $d_0$}
  \end{subfigure}

  \vspace{1mm}

  \begin{subfigure}{0.30\textwidth}
    \centering
    \includegraphics[width=\linewidth]{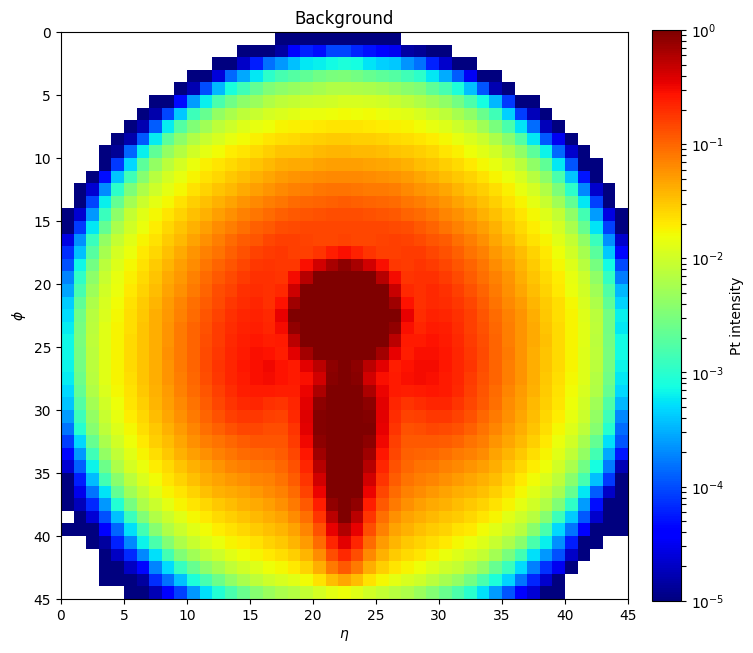}
    \caption{Background, $p_T$}
  \end{subfigure}
  \hspace{0.015\textwidth}
  \begin{subfigure}{0.30\textwidth}
    \centering
    \includegraphics[width=\linewidth]{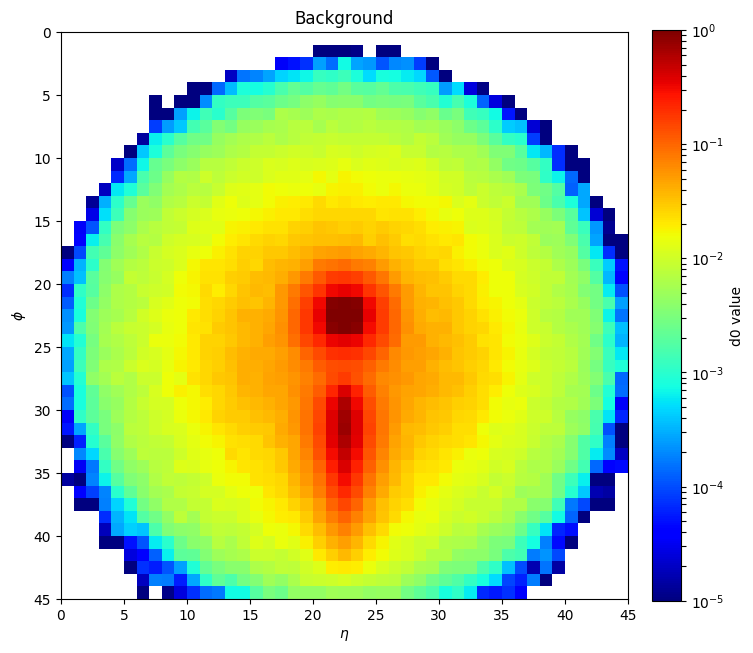}
    \caption{Background, $d_0$}
  \end{subfigure}

  \caption{
  Average jet images for signal $t\to cH$ and background $t\to bW$ events
  in the $p_T$ and $d_0$ channels. All images are aligned and normalized
  according to the preprocessing procedure described in the text.
  }
  \label{fig:jet_images}
\end{figure*}

After applying the event selection criteria described in the previous section, each candidate jet is represented as a multi-channel jet image. The average jet images for signal $t\to cH$ and background $t\to bW$ events are shown in Fig.~\ref{fig:jet_images}. The two displayed channels correspond to the transverse-momentum flow and the displaced-track observable $d_0$, respectively. Both signal and background jets exhibit similar overall radiation patterns due to their boosted multi-prong structure. Their differences are even less apparent in the averaged images, where event-by-event fluctuations and localized substructure features are largely smoothed out.

\begin{figure*}[!htbp]
  \centering
  \begin{subfigure}{0.30\textwidth}
    \centering
    \includegraphics[width=\linewidth]{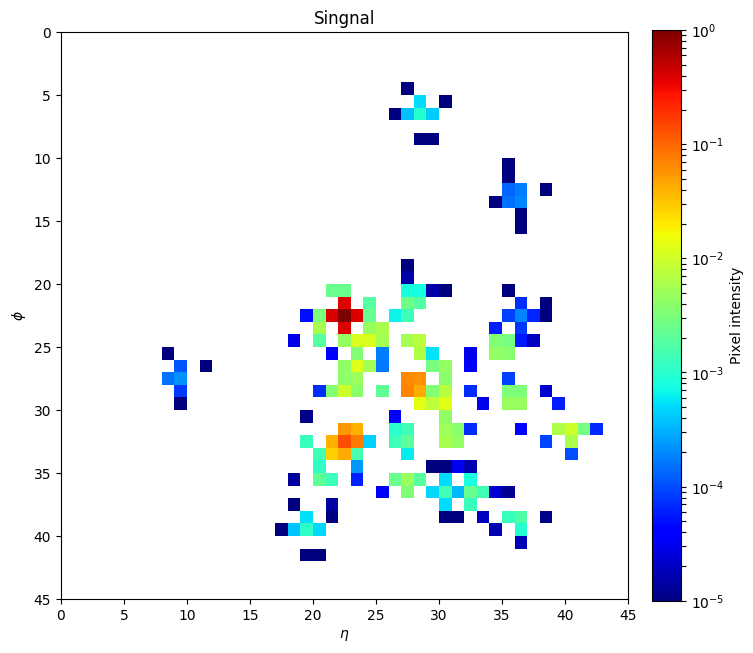}
    \caption{Signal, $p_T$}
  \end{subfigure}
  \hspace{0.015\textwidth}
  \begin{subfigure}{0.30\textwidth}
    \centering
    \includegraphics[width=\linewidth]{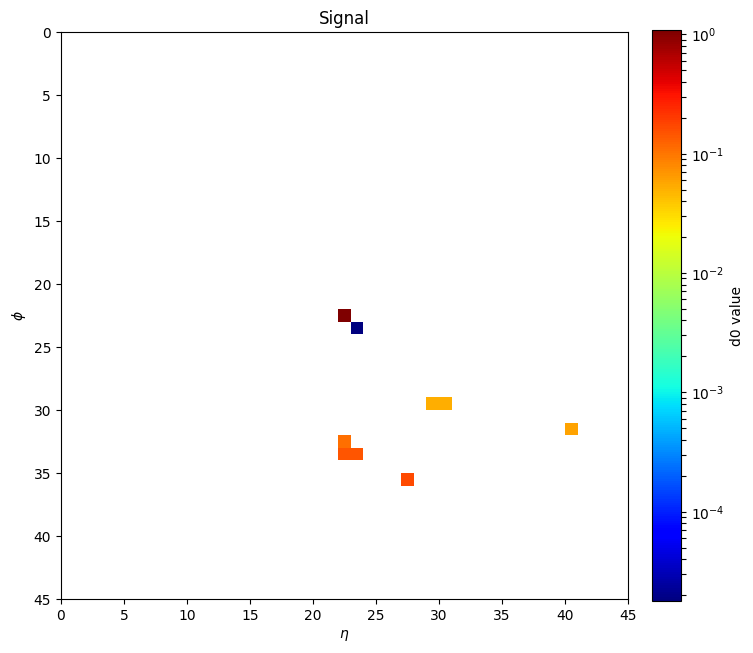}
    \caption{Signal, $d_0$}
  \end{subfigure}

  \vspace{1mm}

  \begin{subfigure}{0.30\textwidth}
    \centering
    \includegraphics[width=\linewidth]{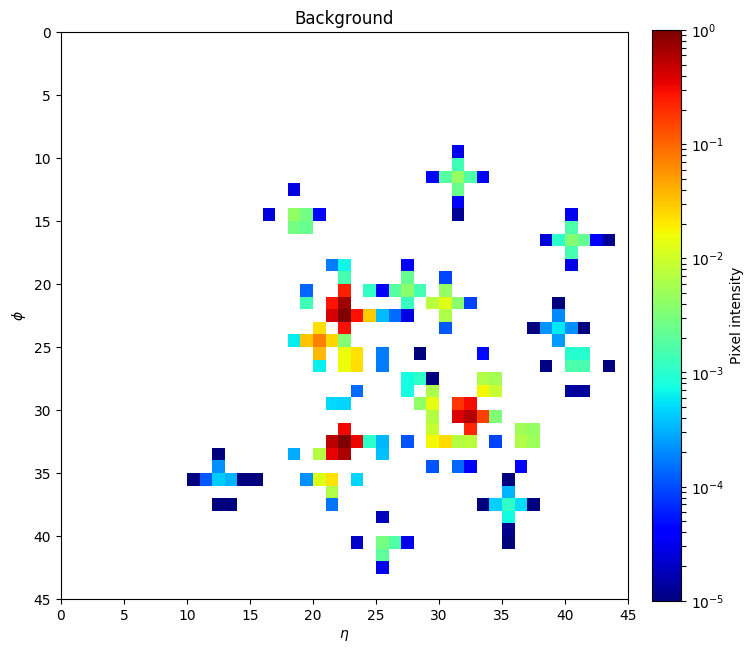}
    \caption{Background, $p_T$}
  \end{subfigure}
  \hspace{0.015\textwidth}
  \begin{subfigure}{0.30\textwidth}
    \centering
    \includegraphics[width=\linewidth]{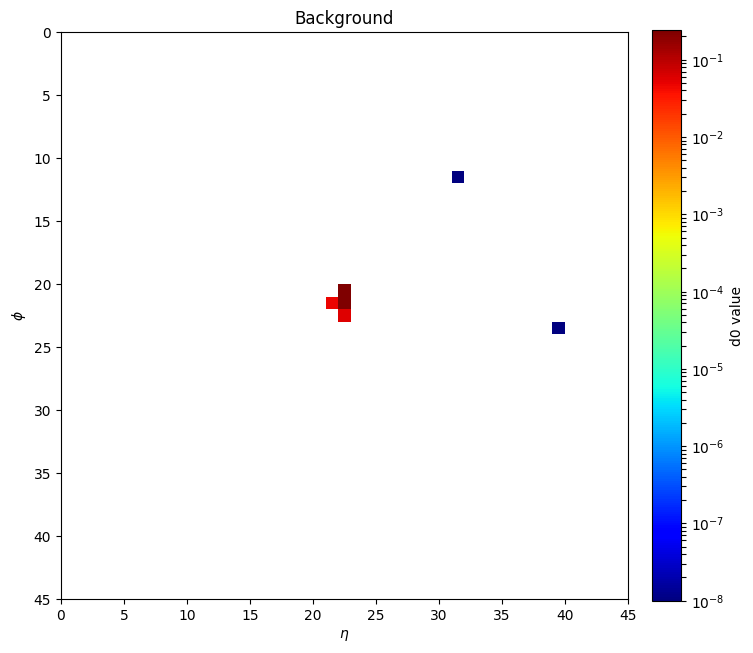}
    \caption{Background, $d_0$}
  \end{subfigure}

  \caption{
  Typical individual jet images for the signal ($t\to cH$) and background
  ($t\to bW$) events after event selection. The top row shows signal events,
  while the bottom row shows background events. The two columns correspond to
  the $p_T$ and $d_0$ channels, respectively.
  }
  \label{fig:individual_jet_images}
\end{figure*}

To further illustrate the characteristic jet topology, we also present several
typical individual jet images for signal and background events in
Fig.~\ref{fig:individual_jet_images}. Unlike the averaged images, these
representative examples retain the event-by-event fluctuations and therefore
provide a more realistic picture of the local energy flow inside a single
large-radius jet. For the signal events, the $p_T$ flow exhibits a
three-prong-like structure, corresponding to the charm quark and the two
$b$ quarks from the Higgs-boson decay. A similar multi-prong pattern is also
reflected in the flavor-sensitive $d_0$ channel, which is related to the
displaced tracks from heavy-flavor hadrons. In contrast, for the background
events, the energy deposition is mainly associated with the heavy-flavor
components, especially the $b$ quark from the top-quark decay and the charm
quark from the hadronic $W$ decay. The corresponding $d_0$ response is therefore
also more localized around these heavy-flavor prongs. These individual images show how the relevant substructure features appear at
the event level and provide a qualitative interpretation of the information
used by the image-based classifier.

Assuming an integrated luminosity of $4~\mathrm{ab}^{-1}$ at
$\sqrt{s}=1.5~\mathrm{TeV}$, we estimate the expected
sensitivity with
\[
Z=\frac{S}{\sqrt{S+B}}.
\]
For the selected classifier threshold, the signal and background efficiencies
are found to be
\[
\varepsilon_S^{\rm CNN}=0.5103,\qquad
\varepsilon_B^{\rm CNN}=8.303\times10^{-4}.
\]

\begin{figure*}[!htbp]
  \centering

  \begin{subfigure}[t]{0.44\textwidth}
    \centering
    \includegraphics[
      width=\linewidth
    ]{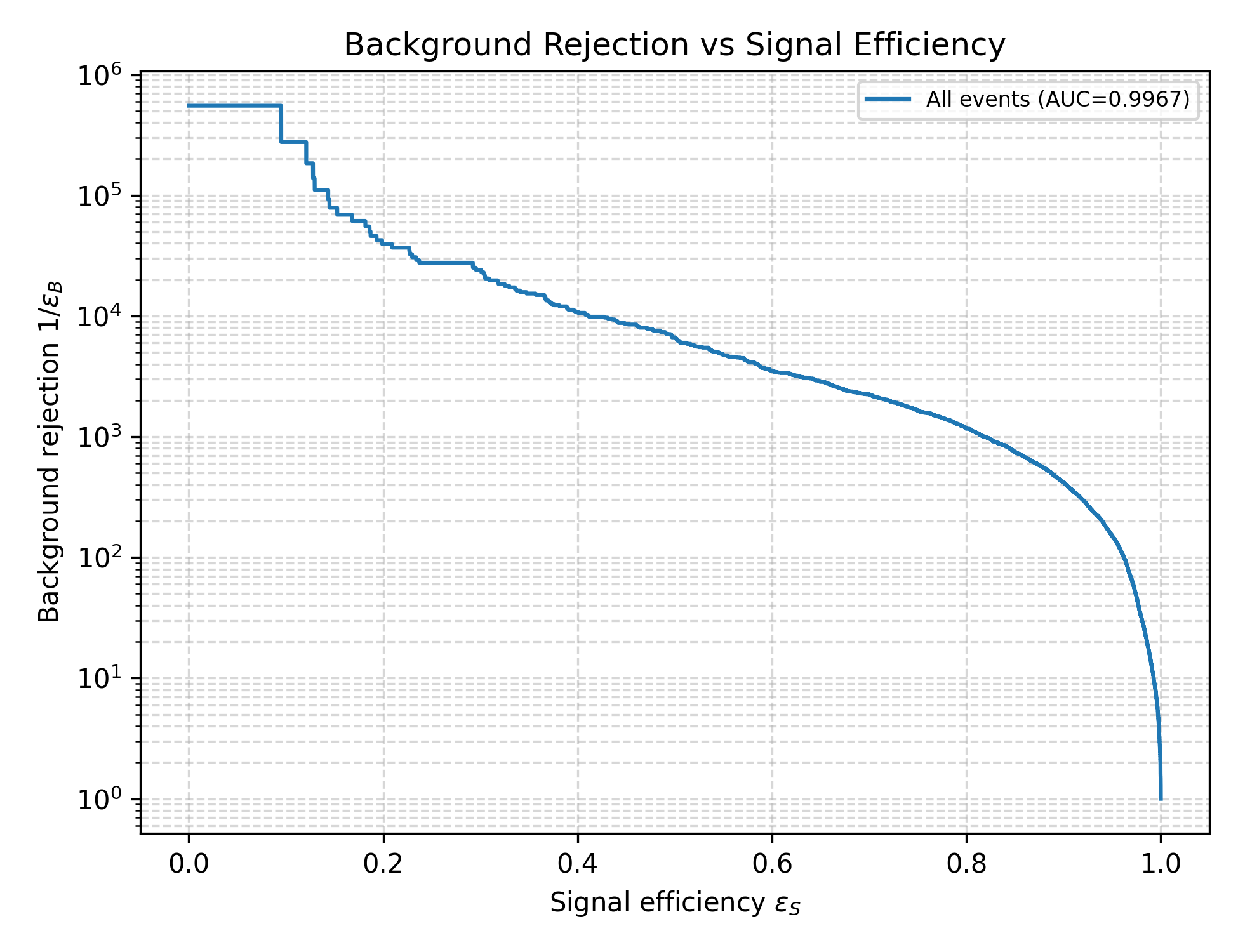}
    \caption{}
    \label{fig:bkg_rejection_sig_eff_ch}
  \end{subfigure}
  \hfil
  \begin{subfigure}[t]{0.44\textwidth}
    \centering
    \includegraphics[
      width=\linewidth,
      trim=0 30 0 0,
      clip
    ]{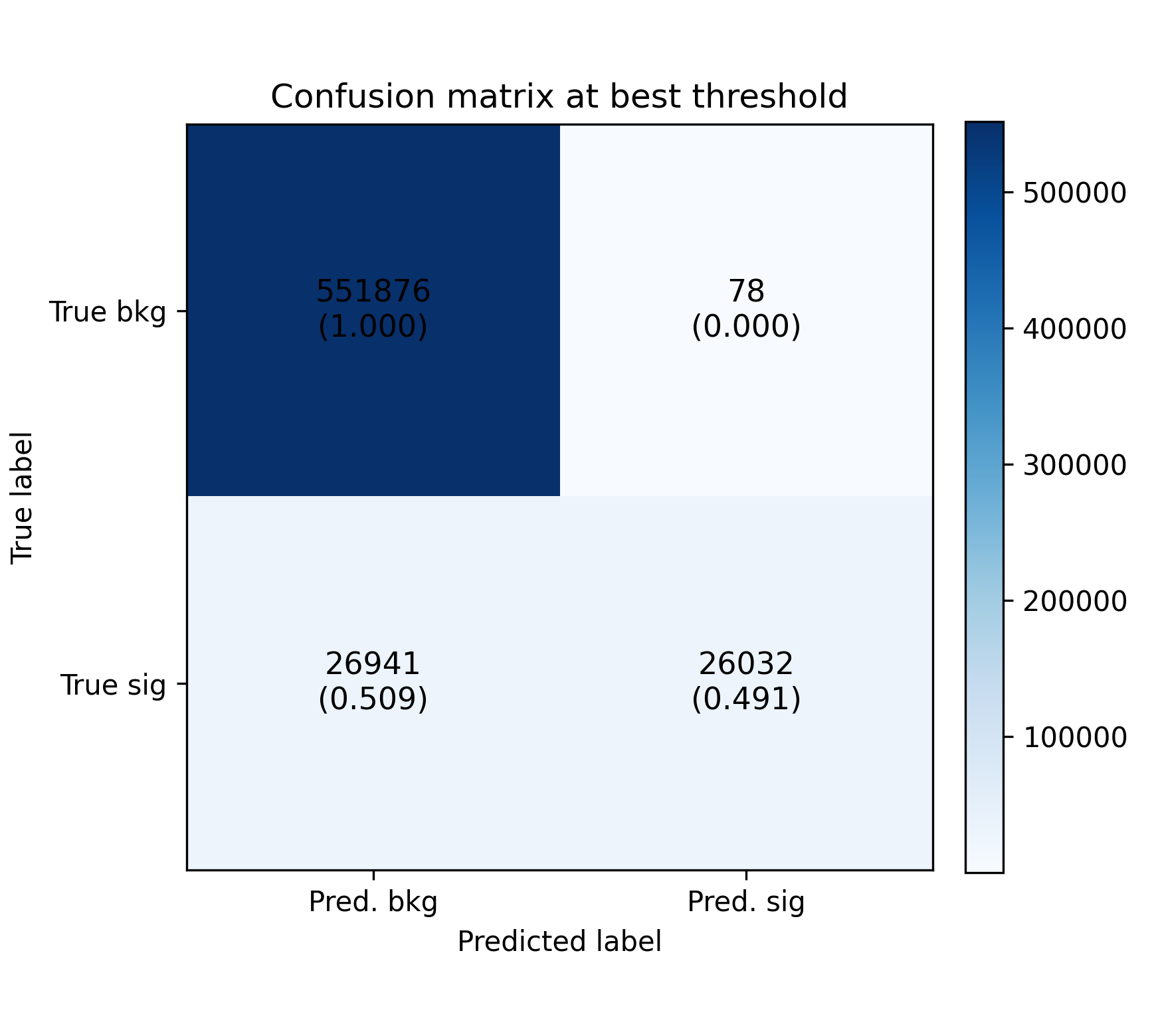}
    \caption{}
    \label{fig:confusion_matrix_ch}
  \end{subfigure}

  \caption{
  Classification performance of the CNN for the $t\to cH$ analysis.
  Left: background rejection $1/\varepsilon_B$ as a function of the signal
  efficiency $\varepsilon_S$.
  Right: confusion matrix evaluated at the operating point threshold.
  }
  \label{fig:cnn_performance_ch}
\end{figure*}

The corresponding classification performance is shown in
Fig.~\ref{fig:cnn_performance_ch}. The background rejection curve illustrates
the operating point used in the sensitivity estimate, while the confusion
matrix shows the event classification result at the same threshold.

Combining them with the preselection efficiencies,
$\varepsilon_S^{\rm pre}=0.53$ and $\varepsilon_B^{\rm pre}=0.552$, we obtain
the total efficiencies
\[
\varepsilon_S=0.2605,\qquad
\varepsilon_B=7.801\times10^{-5}.
\]

Using this operating point, the corresponding expected limit is
\[
\mathrm{BR}(t\to cH)<9.09\times10^{-5},
\]
where $\mathrm{BR}(H\to b\bar b)=0.58$ has been used. Equivalently, the limit
on the product branching ratio is
\[
\mathrm{BR}(t\to cH)\times \mathrm{BR}(H\to b\bar b)
<5.27\times10^{-5}.
\]

\begin{figure}[!htbp]
  \centering
  \includegraphics[width=0.70\linewidth]{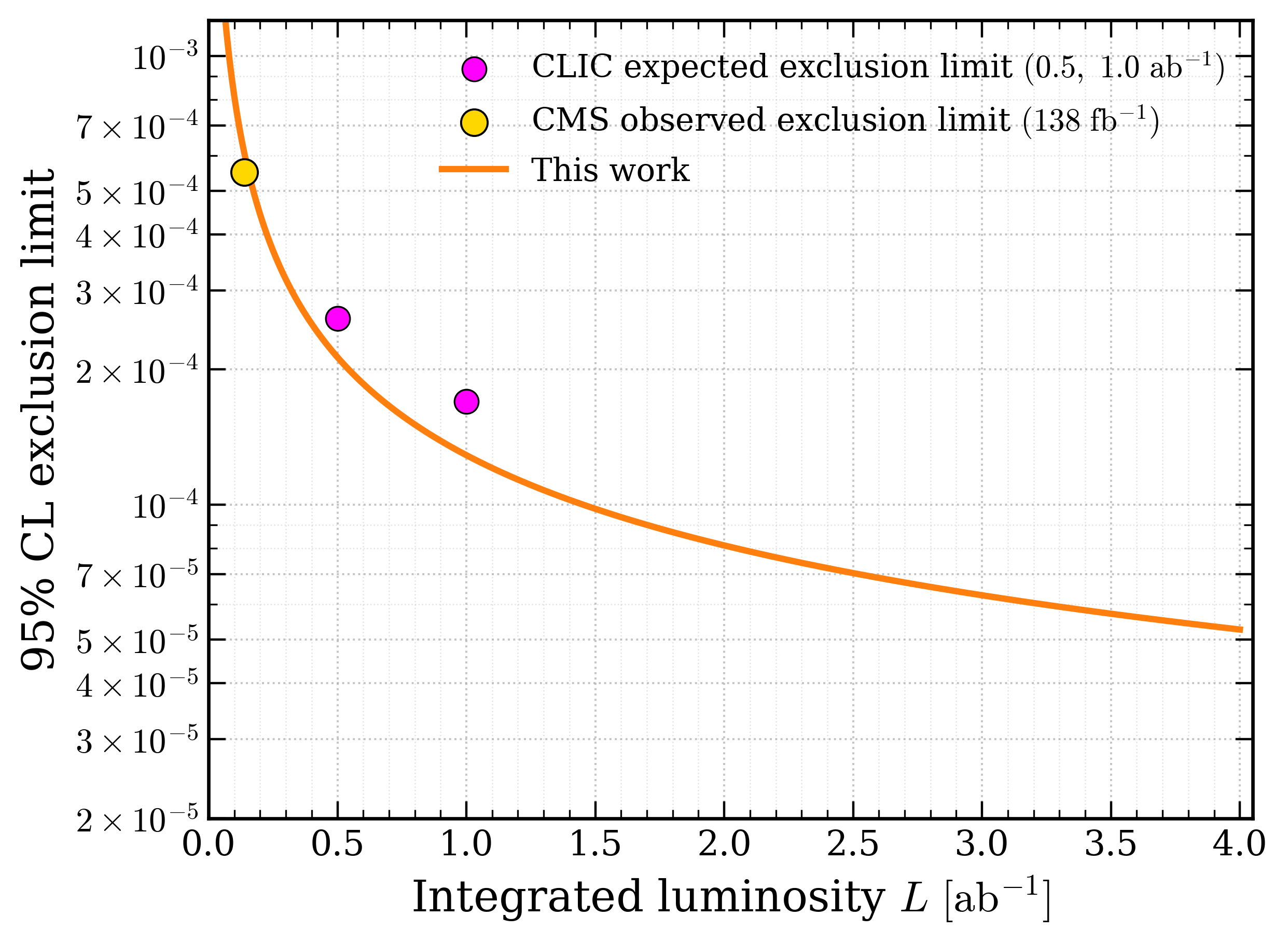}
  \caption{
    The expected $95\%$ C.L. upper limit on
    $\mathrm{BR}(t\to cH)\times \mathrm{BR}(H\to b\bar b)$ as a function of
    the integrated luminosity. The CLIC bound is obtained in this work for
    $\sqrt{s}=1.5~\mathrm{TeV}$. The CLIC expected full-simulation exclusion limit at
    $\sqrt{s}=380~\mathrm{GeV}$, based on $L=0.5~\mathrm{ab}^{-1}$ and
    $L=1~\mathrm{ab}^{-1}$, are also shown for comparison~\cite{Zarnecki:2017cmf}.
    The CMS observed exclusion limit from a search for $t\to cH$, with $H\to b\bar b$,
    produced in $t\bar t$ events at $\sqrt{s}=13~\mathrm{TeV}$ and based on
    $L=138~\mathrm{fb}^{-1}$ is included as an LHC reference~\cite{CMS:2024ubt}.
    }
  \label{fig:exclusion_limit_ch_joint_bb_refs}
\end{figure}

The dependence of the expected product-branching-ratio limit on the integrated
luminosity is shown in Fig.~\ref{fig:exclusion_limit_ch_joint_bb_refs}. The
result obtained in this work is presented together with the CLIC expected
full-simulation results and the CMS expected limit in the $H\to b\bar b$
channel, providing a direct comparison with existing reference sensitivities
for the $t\to cH$ search.

\subsection{$t\to cS$ analysis}

Having studied the $t\to cH$ FCNC signal, we now turn to the
$t\to cS$ channel, where $S$ denotes a new light scalar decaying into
$b\bar b$. To illustrate the sensitivity in different regions of the
light-scalar mass parameter space, We first discuss two representative benchmark masses, $m_S=30$ and
$60~\mathrm{GeV}$, and then summarize the expected limits for the full
mass range $m_S=30$--$80~\mathrm{GeV}$. The expected sensitivity for the
$t\to cS$ process is better than that for $t\to cH$, mainly because the
reconstructed mass of the scalar $S$ can be distinguished from the $W$-boson
mass in the dominant top-pair background, especially for smaller $m_S$.

Since the jet-image construction and preprocessing procedure are the same as
those used in the $t\to cH$ analysis, we do not show the average jet images or
representative individual jet images again. Instead, we summarize the
classifier working points and the resulting expected limits on the product
branching ratio $\mathrm{BR}(t\to cS)\times \mathrm{BR}(S\to b\bar b)$ for these representative benchmark points.

Using the same sensitivity estimate as in the $t\to cH$ analysis, the selected
working points for the two representative masses, $m_S=30$ and
$60~\mathrm{GeV}$, are
\[
\begin{aligned}
\varepsilon_S^{\rm CNN} &= 0.52,\qquad
\varepsilon_B^{\rm CNN} = 4.71\times10^{-5},\\
\varepsilon_S^{\rm CNN} &= 0.3634,\qquad
\varepsilon_B^{\rm CNN} = 4.16\times10^{-5},
\end{aligned}
\]
respectively.

\begin{figure}[!htbp]
  \centering
  \includegraphics[width=0.70\linewidth]{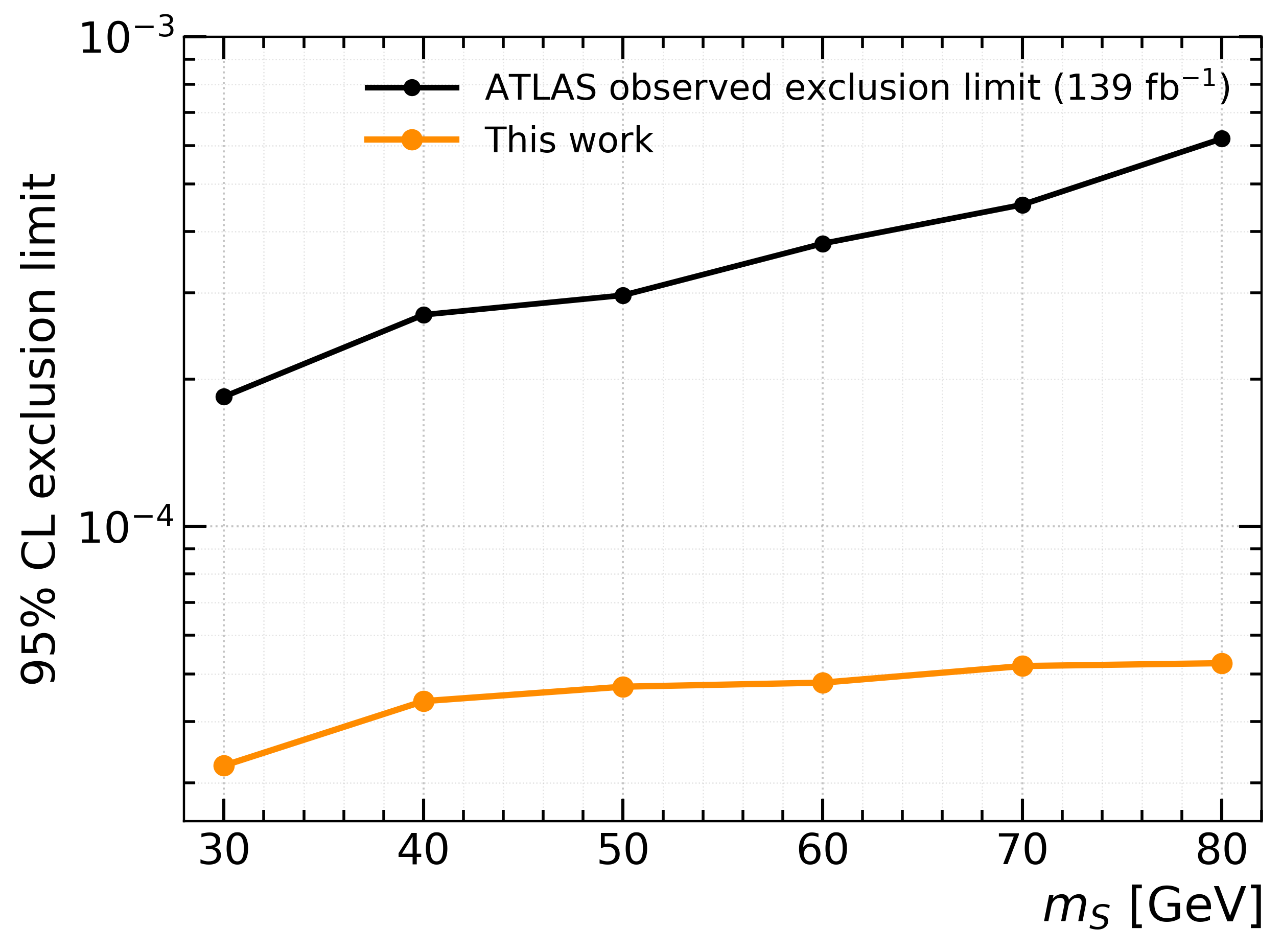}
  \caption{
   The expected $95\%$ C.L. upper bounds on
   $\mathrm{BR}(t\to cS)\times \mathrm{BR}(S\to b\bar b)$ in the
   $b\bar b$ channel. The CLIC bounds are obtained for
   $\sqrt{s}=1.5~\mathrm{TeV}$ and $L=4~\mathrm{ab}^{-1}$. The ATLAS observed exclusion limit
   from a search for $t\to qX$, with $X\to b\bar b$, produced in
   $t\bar t$ at $\sqrt{s}=13~\mathrm{TeV}$ and based on $139~\mathrm{fb}^{-1}$ of data is also shown as the
   black solid line\cite{ATLAS:2023mcc}.
   }
  \label{fig:compare_clic_atlas_tcs_30_80GeV_simple}
\end{figure}

These efficiencies are evaluated with respect to the preselected candidate-jet
samples. After combining them with the preselection efficiencies, the total
efficiencies entering the event-yield estimate are
\[
\begin{aligned}
\varepsilon_S &= 0.2827\qquad
\varepsilon_B = 2.6\times10^{-5},\\
\varepsilon_S &= 0.1982,\qquad
\varepsilon_B= 2.3\times10^{-5},
\end{aligned}
\]
for $m_S=30$ and $60~\mathrm{GeV}$, respectively.

At an integrated luminosity of $4~\mathrm{ab}^{-1}$, the expected $95\%$ C.L.
upper limits are
\[
\begin{aligned}
\mathrm{BR}(t\to cS)\times \mathrm{BR}(S\to b\bar b)
&<3.25\times10^{-5},\\
\mathrm{BR}(t\to cS)\times \mathrm{BR}(S\to b\bar b)
&<4.80\times10^{-5},
\end{aligned}
\]
for $m_S=30$ and $60~\mathrm{GeV}$, respectively.

The expected upper limit on
$\mathrm{BR}(t\to cS)\times \mathrm{BR}(S\to b\bar b)$ are summarized in
Fig.~\ref{fig:compare_clic_atlas_tcs_30_80GeV_simple} as a function of the
scalar mass. For comparison, the ATLAS observed exclusion limit from a search for
$t\to qX$, with $X\to b\bar b$, produced in $t\bar t$ events is also shown.

\section{Conclusions}
Top-quark FCNC decays provide a particularly sensitive probe of physics beyond
the Standard Model. In this work, we have studied the top-quark FCNC decay
channels $t\to cH$ and $t\to cS$ in top-pair production at CLIC with
$\sqrt{s}=1.5~\mathrm{TeV}$. We focus on a distinct kinematic regime where the top quarks are produced in highly boosted configuration. A large-radius $R=1.0$ is employed to capture the
collimated hadronic decay products of the boosted top quark, and multi-channel
jet images are constructed from the jet constituents. We then use convolutional
neural networks (CNNs), which are well suited to exploit the local spatial
correlations present in these images, to discriminate the FCNC signals from the
dominant Standard Model background arising from hadronic $t\to Wb$ decays.

Based on the analysis, for the Higgs boson channel we obtain the expected
$95\%$ C.L. upper limit
\[
\mathrm{BR}(t\to cH)\times \mathrm{BR}(H\to b\bar b)
<5.27\times10^{-5}
\]
with an integrated luminosity of $4~\mathrm{ab}^{-1}$.

For the case of a new light scalar $S$, we present the results in terms of the
product branching ratio
$\mathrm{BR}(t\to cS)\times \mathrm{BR}(S\to b\bar b)$, so that they can be
interpreted for a broader class of light-scalar scenarios. We also present the the exclusion limit curve for $m_S$ value ranging from 30 GeV to 60 GeV with an luminosity of $4~\mathrm{ab}^{-1}$. Taking two representative benchmark masses, $m_S=30$ and $60~\mathrm{GeV}$, the
corresponding expected upper limits on the product branching ratio are
\[
\begin{aligned}
\mathrm{BR}(t\to cS)\times \mathrm{BR}(S\to b\bar b)
&<3.25\times10^{-5}, \qquad m_S=30~\mathrm{GeV},\\
\mathrm{BR}(t\to cS)\times \mathrm{BR}(S\to b\bar b)
&<4.80\times10^{-5}, \qquad m_S=60~\mathrm{GeV}.
\end{aligned}
\]

In summary, our results demonstrate that rare top-quark decays at future
high-energy CLIC collider provide a promising probe of new physics. The CNN architecture based on
multi-channel jet images offers significant potential for improving the
sensitivity to FCNC top-quark decay signatures, not only for the Higgs
channel but also for a broader class of light-scalar scenarios. This approach
can be further extended to other rare top processes and exotic decay scenarios.

\section*{Acknowledgements}
We would like to thank Jinmian Li and Rao Zhang for helpful discussions and insightful comments on the draft. In addition, we would like to acknowledge Yi-Qi Wang for her contributions in the early stages of this work. This work was supported in part by the National Natural Science Foundation of China under Grants No. 12575106 and 12147214. SY was supported in part by the Basic Scientific Research Project of the Education Department of Liaoning Province under Grant No.
LJ2610165. JY was supported in part by the Basic Scientific Research Project of the Education Department of Liaoning Province under Grant No. LJ212510165024.

\bibliography{references}

\end{document}